%Paper: astro-ph/9504022
%From: mannheim@main.phys.uconn.edu (Philip Mannheim)
%Date: Fri, 7 Apr 95 08:22:17 EDT

\magnification=1000
 \baselineskip=0.50truecm
%\baselineskip=0.82truecm
\centerline {\bf LINEAR POTENTIALS IN GALAXIES AND CLUSTERS OF GALAXIES}
\bigskip
\centerline {\bf Philip D. Mannheim}
\centerline {Department of Physics}
\centerline {University of Connecticut}
\centerline {Storrs, CT 06269-3046}
\smallskip
\centerline{mannheim@uconnvm.uconn.edu}
\bigskip
\centerline {\bf Abstract}
\medskip
In a previous paper we presented a typical set of galactic rotation curves
associated with the linear gravitational potential of the conformal invariant
fourth order theory of gravity which has recently been advanced by Mannheim
and Kazanas as a candidate alternative to the standard second order
Newton-Einstein theory. Reasonable agreement with data was obtained for four
representative galaxies without the need for any non-luminous or dark matter.
In this paper we present the associated formalism and compare and contrast the
linear potential explanation of the general systematics of galactic rotation
curves and the associated Tully-Fisher relation with that of the standard dark
matter theory. Additionally, we show that the conformal gravity picture appears
to have survived the recent round of microlensing observations unscathed.
Finally, we make a first application of the conformal theory to the larger
distance scale associated with a cluster of galaxies, with the theory being
found to give a reasonable value for the mean velocity of the virialized core
of the typical Coma cluster, again without the need to invoke dark matter.
\vskip 4.00truecm
\noindent
March,
1995$~~~~~~~~~~~~~~~~~~~~~~~~~~~~~~~~~~~~~~~~~~~~~~~~~~~~~~~~~~~~~~~~~~~~~~~~
{}~~~~~~~~~~~~~~$UCONN-95-2
\vfill\eject
\hoffset=0.0truein
\hsize=6.5 truein
\noindent
{\bf (1) Introduction}
\medskip
At the present time there is little doubt in the general community as to the
correctness of the standard second order Newton-Einstein theory of gravity.
However given the fact that the application of this theory to currently
available
astrophysical and cosmological data obliges the Universe to be composed
of overwhelming amounts of non-luminous or dark matter, it is a well
established scientific tradition to pause and question so startling an
implication, and to at
least consider the possibility that this need for dark matter might instead
actually be
signaling a possible breakdown of the standard theory. Since this apparent
need for dark matter is manifest on essentially every single distance scale
from galactic all the way up to cosmological, while no such need is generally
manifest on the much shorter distance scales where the standard gravitational
theory was
originally established in the first place, it is thus natural to consider the
possibility that new physics (one might even refer to it as dark physics) may
be opening up on these bigger distance scales. Indeed, rather than
interpreting essentially every single current large distance scale
gravitational observation as
yet further evidence for the existence of dark matter (the common practice in
both the learned
and the popular literature), these selfsame data can just as equally be
regarded as signaling the repeated failure of the standard theory;
and definitively so if the bulk of the matter which actually
exists in the Universe is in fact luminously observable. Thus the
psychologically unwelcome
empirical possibility suggested by the data (and now all the more so given the
apparent failure
so far of the current round of direct microlensing and optical dark and faint
matter
searches to validate the standard galactic spherical dark halo scenario) is
that Newton's Law of
Gravity may not be the correct weak gravitational theory on large distance
scales, and that,
accordingly, second order Einstein gravity may not then be the correct
covariant one.

Now of course both the Newton and Einstein theories enjoy many successes
(enough to convince most people that they are no longer even challengeable at
all), and thus any alternate theory of gravity must be able to recover all
their established features. To achieve this, one way to proceed is to begin
with galactic rotation curve data (perhaps the most clear cut and well explored
situation
where the Newton-Einstein theory demands dark matter) and try to extract out
a new weak gravity limit which encompasses Newton in an appropriate limit
(see e.g. Milgrom 1983a, b, c and Sanders 1990) with a view to then
subsequently
working upwards to a covariant generalization (a program which is still in
progress
- see Bekenstein 1987 and Sanders 1990 for recent reviews). However, in order
to ensure
encompassing the Einstein successes from the outset, there is also much merit
in beginning
covariantly and then working downwards to a weak gravity limit (an approach
which then
comes with the additional challenge of not knowing what weak gravity limit may
eventually
ensue until after solutions to any candidate alternate covariant theory have
actually
been found), with this latter approach having actually been advanced and
explored
in the literature by Mannheim and Kazanas in a recent
series of papers (Mannheim and Kazanas 1989, 1991, 1994, Mannheim 1990,
1992, 1993a, 1993b, 1994, Kazanas 1991, Kazanas and Mannheim 1991a, 1991b).
Recognizing that there is currently no known theoretical reason which would
select out the standard second order Einstein theory from amongst the infinite
class of (all order) covariant, pure metric based theories of gravity that one
could in principle at least consider, Mannheim and Kazanas reopened the
question of what the correct covariant theory might then be and developed an
approach which
fixes the gravitational action by imposing an additional fundamental
principle above and beyond covariance, namely that of local scale or conformal
invariance, i.e. invariance under any and all local conformal stretchings
$g_{\mu\nu}(x) \rightarrow \Omega^2(x) g_{\mu\nu}(x)$ of the geometry, this
being
the invariance which is now believed to be possessed by the other three
fundamental interactions, the strong, the electromagnetic and the weak. This
invariance forces gravity to then be described uniquely by the fourth order
action
$$I_W = -\alpha \int d^4x (-g)^{1/2}
C_{\lambda\mu\nu\kappa}C^{\lambda\mu \nu\kappa}
=-2\alpha \int d^4 x (-g)^{1/2}
(R_{\lambda\mu}R^{\lambda\mu} - (R^{\alpha}_{\phantom {\alpha}\alpha})^{2}/3)
\eqno(1) $$
where $C_{\lambda\mu\nu\kappa}$ is the conformal Weyl tensor and $\alpha$
is a purely dimensionless coefficient. In their original paper Mannheim
and Kazanas (1989) obtained the complete and exact, non-perturbative exterior
vacuum solution associated with a static,
spherically symmetric gravitational source such as a star in this theory, viz.
$$-g_{00}= 1/g_{rr}=1-\beta(2-3 \beta \gamma )/r - 3 \beta \gamma
+ \gamma r - kr^2 \eqno(2)$$
(for line element $ds^2=-g_{00}c^2dt^2-g_{rr}dr^2-r^2d\Omega$)
where $\beta, \gamma,$ and $k$ are three appropriate dimensionful
integration constants. As can be seen, for small enough values of the linear
and quadratic terms (i.e. on small enough distance scales) the solution
reduces to the familiar Schwarzschild solution of Einstein gravity, with the
conformal theory then enjoying the same static successes as the Einstein theory
on
those distance scales. On larger distance scales, however, the theory begins to
differ
from the Einstein theory through the linear potential term, and (with the
quadratic term only possibly being important cosmologically, and with both the
$\beta\gamma$ product terms being found to be numerically negligible in the
fits of Mannheim 1993b) then yields a non-relativistic gravitational
potential
$$V(r)=-\beta c^2/r+ \gamma c^2r/2 \eqno(3)$$
which may be fitted to data whenever the weak gravity limit is applicable.

The conformal theory thus not only generalizes Newton (Eq.
(3)) it also generalizes Schwarzschild (Eq. (2)), and even does so in way
which is then able to naturally recover both the Newton and Schwarzschild
phenomenologies on the appropriate distance scales. Since the conformal theory
recovers the requisite solutions to Einstein gravity on small enough distance
scales (even while never recovering the Einstein Equations themselves -
observation only demands the recovery of the solutions not of the equations),
that fact alone makes the theory indistinguishable from and just as viable as
the Einstein theory on those distance scales, something recognized by
Eddington (1922) as far back as the very early days of Relativity. (Eddington
was not
aware of the full exact solution of Eq. (2) but was aware that it was a
solution
to fourth order gravity in the restricted case where $\gamma=0$. It was only
much
later that the complete and exact solution of Eq. (2) was found and that its
consistency
was established by successfully matching it on to the associated
exact interior solution (Mannheim and Kazanas 1994)). Thus in this sense
conformal gravity should always have been considered as a viable explanation
of solar system physics. That it never was so considered was in part due to
the fact that strict conformal symmetry requires that all particles be
massless, something which would appear to immediately rule the symmetry out.
However,
with the advent of modern spontaneously broken gauge theories manifest in the
other three fundamental interactions, it is now apparent that mass can still
be generated in the vacuum in otherwise dimensionless theories like the one
associated with the action of Eq. (1). (And, interestingly, such dynamical mass
generation is even found to still lead to geodesic motion (Mannheim 1993a),
despite the fact that the associated mass generating Higgs scalar field
which accompanies a test particle carries its own energy and momentum which the
gravitational field also sees). Hence, it would appear that today the only
non-relativistic way to distinguish between the two covariant theories is to
explore
their observational implications on larger distance scales
where the linear potential term first makes itself manifest.

A first step towards this phenomenological end was taken recently by Mannheim
(1993b) with the above non-relativistic potential $V(r)$ being used in
conjunction
solely with observed surface brightness data (i.e. without assuming any dark
matter)
to fit the rotation
curves of four representative galaxies. The particular choice of galaxies was
guided by
the recent comprehensive survey of the $HI$ rotation curves of spiral galaxies
made by
Casertano and van Gorkom (1991) who found that those data fall into essentially
four general groups characterized by specific correlations between the maximum
rotation
velocity and the luminosity. In order of increasing luminosity the four groups
are
dwarf, intermediate, compact bright, and large bright galaxies. Thus one
representative
galaxy from each group was studied, respectively the galaxies DDO 154 (a gas
dominated
rather than star dominated galaxy), NGC 3198, NGC 2903, and NGC 5907, with the
fitting of
Mannheim (1993b) being reproduced here as Fig. (1). (The reader is referred to
the
original paper for details). For NGC 3198 the rotation curve of Begeman (1989)
and the
surface brightness data of Wevers et al. (1986) and Kent (1987) were used, for
NGC 2903
the data were taken from Begeman (1987) and Wevers et al. (1986), for NGC 5907
from
van Albada and Sancisi (1986) and Barnaby and Thronson (1992), and for DDO 154
from
Carignan and Freeman (1988) and Carignan and Beaulieu (1989). (While Carignan
and his
coworkers favor a distance of 4 Mpc to DDO 154, Krumm and Burstein 1984 favor
10 Mpc. Since the gas contribution is extremely distance sensitive, for
completeness we
opted to fit this galaxy at both the candidate distances). As can be seen from
Fig. (1),
the conformal theory appears to be able to do justice to a data set which
involves a broad
range of luminosities, and to even do so without the need for dark matter, a
point we
analyze further below.

In order to apply the linear potential to an extended object such as a disk
it was found helpful to develop a general formalism, with the results of the
formalism
being used in Mannheim (1993b) to produce the fits of Fig. (1). In Sec. (2) of
the present paper we
present the actual details of the derivation of the formalism (something that
will be
useful for future studies), with the formalism actually even being of interest
in its
own right since it extends to linear potentials the earlier work of Toomre
(1963), Freeman (1970),
and Casertano (1983) on Newtonian disks. In Sec. (3) we analyze some of the
general systematics of
galactic rotation curve fitting, and compare and contrast our fitting with that
of the standard dark
matter theory. Additionally in Sec. (3) we discuss the status of the
Tully-Fisher relation (Tully
and Fisher 1977) in conformal gravity, a relation which informs us that for
regular spiral galaxies
the velocity (i.e. the total gravitational potential) is universally normalized
to the luminosity, a
quantity which is fixed by the explicit amount of visible matter (i.e. that
matter which is not
non-luminous) in each of the relevant galaxies. The very existence of the
Tully-Fisher relation thus
quite strongly suggests that the total gravitational potential is fixed by the
amount of luminous
matter in a galaxy, and that galaxies should accordingly be predominantly
luminous. Consequently,
given the rotation curve data, it would appear that it is the rule for
calculating the potential
which therefore needs to be changed, to thus necessitate deviations from Newton
on galactic distance
scales such as those predicted in our theory. Now a theory	with a linearly
rising potential will
lead to ever bigger deviations from Newton on ever bigger distance scales, a
trend which in a
first approximation nicely parallels that found in the standard theory where
ever increasing amounts
of dark matter are required as distance scales get larger. Consequently, in
Sec. (4) we both
develop an appropriate formalism for and make a first application of the
conformal theory to the
first available scale beyond galaxies, namely that of clusters of galaxies.
Interestingly, we find
that the conformal theory deviates from Newton there by just the amount needed
to nicely accommodate
a virialized cluster core without invoking dark matter, though the theory may
turn out to have some
difficulties should entire clusters prove to be virialized. In Sec. (5) we
discuss the implications
for gravitational theory of the recent round of microlensing searches for
astrophysical dark matter,
and show that the data presented so far leave the conformal gravity theory
viable.

\medskip
\noindent
{\bf (2) The Potential of an Extended Disk}
\medskip

In order to handle the weak gravity potential of an extended object such as a
disk
of stars each with gravitational potential $V(r)=-\beta c^2/r+ \gamma c^2r/2 $
many
ways are possible, with perhaps the most popular being a method due to Toomre
(1963)
which was originally developed for thin Newtonian disks. Since that method does
not
immediately appear to generalize to linear potentials, we have instead
generalized his
approach first to non-thin Newtonian disks (a step also taken by  Casertano
1983) and
then to disks with linear potentials.
To determine the Newtonian potential of an axially symmetric
(but not yet necessarily thin) distribution of matter sources with
matter volume density function
$\rho(R,z^{\prime})$ we need to evaluate the quantity
$$V_{\beta}(r,z)=-\beta c^2
\int_0^{\infty}dR \int_0^{2\pi}d\phi^{\prime}
\int_{-\infty}^{\infty}dz^{\prime}
{R\rho(R,z^{\prime}) \over
(r^2+R^2-2rRcos\phi^{\prime}+(z-z^{\prime})^2)^{1/2}} \eqno(4)$$
where $R,~\phi^{\prime},~z^{\prime}$ are cylindrical source coordinates
and $r$ and $z$ are the only observation point coordinates of relevance.
To evaluate Eq. (4) it is convenient to make use of the cylindrical
Green's function Bessel function expansion
$${1 \over \vert {\bf r} -{\bf r^{\prime}} \vert }=\sum_{m=-\infty}^{\infty}
\int_0^\infty dk J_m(kr)J_m(kr^{\prime})
e^{im(\phi-\phi^{\prime})-k \vert  z -  z^{\prime} \vert } \eqno(5)$$
whose validity can readily be checked by noting that use of the identity
$$\nabla^2[J_m(kr)e^{im\phi-k \vert  z -  z^{\prime} \vert }]
=-2kJ_m(kr)e^{im\phi}\delta(z-z^{\prime}) \eqno(6)$$
leads to the relation
$$\nabla^2 \left( {1 \over \vert {\bf r} -{\bf r^{\prime}} \vert }\right)=
-4\pi\delta^3({\bf r} -{\bf r^{\prime}}) \eqno (7)$$
(In his original study Toomre used a Bessel function
discontinuity formula (essentially Eq. (6))
which only appears to be applicable to thin disks. Using
the full completeness properties of the Bessel functions enables us to treat
non-thin disks as well).
While Eq. (5) is standard, it is not utilized as often as the more
familiar modified Bessel function expansion
$${1 \over \vert {\bf r} -{\bf r^{\prime}} \vert }={2 \over \pi}
\sum_{m=-\infty}^{\infty}
\int_0^\infty dk cos[k (  z -  z^{\prime})]
I_m(kr_{<})K_m(kr_{>}) e^{im(\phi-\phi^{\prime})}\eqno (8)$$
since the product of the two modified Bessel functions has much better
convergence properties at
infinity than the product of the two ordinary Bessel functions. Nonetheless,
the ordinary Bessel
functions do actually vanish at infinity which is sufficient for our purposes
here. A
disadvantage of the expansion of Eq. (8) is that it
involves oscillating $z$ modes rather than the bounded $z$ modes
given in Eq. (5), with the bounded form of Eq. (5) actually being extremely
convenient
for a disk whose matter distribution is concentrated around $z=0$. An
additional
shortcoming of the expansion of Eq. (8) is that when it is inserted into
Eq. (4) it requires the $R$ integration range to be broken up into two
separate pieces at the point of observation.
However, inserting Eq. (5) into Eq. (4) leads to
$$V_{\beta}(r,z)=-2\pi\beta c^2\int_{0}^{\infty} dk\int_{0}^{\infty}dR
\int_{-\infty}^{\infty}dz^{\prime}
R \rho(R,z^{\prime})J_0(kr)J_0(kR) e^{-k\vert z -  z^{\prime}\vert}\eqno(9)$$
which we see requires no such break up. Finally, taking the disk to be
infinitesimally thin (viz. $\rho(R,z^{\prime})=\Sigma(R)\delta(z^{\prime})$)
then yields for points with $z=0$ the potential
$$V_{\beta}(r)=-2\pi\beta c^2\int_{0}^{\infty} dk\int_{0}^{\infty}dR
R \Sigma(R)J_0(kr)J_0(kR) \eqno(10)$$
which we immediately recognize as Toomre's original result for an
infinitesimally thin disk. In passing we note that Eq. (9) also holds for
points
which do not lie in the $z=0$ plane of the disk, and also applies to disks
whose
thickness may not in fact be negligible, with the form of Eq. (9) being
particularly convenient if the fall-off of the matter distribution in the
$z$ direction is itself exponential (see below).

For our purposes here, the expansion of Eq. (5) can immediately be
applied to the linear potential case too, and this leads directly
(on setting $\vert {\bf r} -{\bf r^{\prime}} \vert =
({\bf r} -{\bf r^{\prime}})^2/
\vert {\bf r} -{\bf r^{\prime}} \vert$) to the potential
$$V_{\gamma}(r,z)={\gamma c^2\over 2}
\int_0^{\infty}dR \int_0^{2\pi}d\phi^{\prime}
\int_{-\infty}^{\infty}dz^{\prime}
R\rho(R,z^{\prime})
[r^2+R^2-2rRcos\phi^{\prime}+(z-z^{\prime})^2]^{1/2}$$
$$=\pi\gamma c^2
\int_0^{\infty}dk \int_0^{\infty}dR \int_{-\infty}^{\infty}dz^{\prime}
R\rho(R,z^{\prime})
[(r^2+R^2+(z-z^{\prime})^2)J_0(kr)J_0(kR)$$
$$-2rR J_1(kr)J_1(kR)]
e^{-k\vert z -  z^{\prime}\vert} \eqno(11)$$
Equation (11) then  reduces at $z=0$ for infinitesimally thin disks to the
compact expression
$$V_{\gamma}(r)=\pi\gamma c^2\int_{0}^{\infty} dk\int_{0}^{\infty}dRR
\Sigma(R)[ (r^2+R^2)J_0(kr)J_0(kR) -2rR J_1(kr)J_1(kR)]
\eqno(12)$$

If the $k$ integrations are performed first in Eqs. (10) and (12) they
lead to singular hypergeometric functions whose subsequent $R$
integrations contain infinities which, even while they are in fact mild enough
to be
integrable (as long as $\Sigma(R)$ is sufficiently damped at infinity),
nonetheless require a little care when being carried out numerically.
Thus unlike the sphere whose potential is manifestly finite at every step of
the
calculation, the disk, because of its lower dimensionality, actually encounters
infinities
at any interior point of observation on the way to a final finite answer.
However, since
the final
answer is finite, it should be possible to obtain this answer without ever
encountering
any infinities at any stage of the calculation at all; and indeed, if the
distribution
function $\Sigma(R)$ is available in a closed form, then performing the $R$
integration before the $k$ integration can yield a calculation which is finite
at every
stage. Thus, for the exponential disk
$$\Sigma(R)=\Sigma_0e^{-\alpha R} \eqno (13)$$
where $1/\alpha=R_0$ is the scale length of the disk and $N=2\pi\Sigma_0 R_0^2$
is the number of stars in the disk, use of the standard Bessel
function integral formulas
$$\int_0^\infty dR RJ_0(kR)e^{-\alpha R}={\alpha \over
(\alpha^2+k^2)^{3/2}} \eqno(14)$$
$$\int_0^\infty dk {J_0(kr) \over (\alpha^2+k^2)^{3/2}}
  =(r/2\alpha)[I_0(\alpha r/2)K_1(\alpha r/2)-
I_1(\alpha r/2)K_0(\alpha r/2)]
\eqno(15)$$
then leads directly to Freeman's original result, viz.
$$V_{\beta}(r)=-2\pi\beta c^2\Sigma_0 \int_{0}^{\infty}dk
{\alpha J_0(kr) \over (\alpha^2+k^2)^{3/2}}$$
$$= -\pi\beta c^2\Sigma_0 r[I_0(\alpha r/2)K_1(\alpha r/2)-
I_1(\alpha r/2)K_0(\alpha r/2)]\eqno(16)$$
for the Newtonian potential of an exponential disk. The use of the additional
integral
formula
$$\int_0^\infty dR R^2J_1(kR)e^{-\alpha R}={3\alpha k\over
(\alpha^2+k^2)^{5/2}} \eqno (17)$$
and a little algebra (involving eliminating $J_1(kr)=-(dJ_0(kr)/dk)/r$ via
an integration by parts) enable us to obtain for the
linear potential contribution the expression
$$V_{\gamma}(r)=
\pi \gamma c^2\Sigma_0
\int_{0}^{\infty}dk
\left( {\alpha r^2 J_0(kr) \over (\alpha^2+k^2)^{3/2}}
-{9\alpha J_0(kr) \over (\alpha^2+k^2)^{5/2}}
+{15\alpha^3 J_0(kr) \over (\alpha^2+k^2)^{7/2}}
-{6\alpha kr J_1(kr) \over (\alpha^2+k^2)^{5/2}} \right)$$
$$=\pi\gamma c^2\Sigma_0
\int_{0}^{\infty}dk J_0(kr)
\left( {\alpha r^2 \over (\alpha^2+k^2)^{3/2}}
+{15\alpha \over (\alpha^2+k^2)^{5/2}}
-{15\alpha^3 \over (\alpha^2+k^2)^{7/2}} \right)
\eqno(18)$$
\noindent
Equation (18) is readily evaluated through use of the modified
Bessel function recurrence relations
$$ I_0^{\prime}(z)=I_1(z)~~~,~~~I_1^{\prime}(z)=I_0(z)-I_1(z)/z $$
$$ K_0^{\prime}(z)=-K_1(z)~~~,~~~K_1^{\prime}(z)=-K_0(z)-K_1(z)/z
\eqno(19)$$
in conjunction with Eq. (15) and its derivatives, and yields
$$V_{\gamma}(r)= \pi\gamma c^2\Sigma_0
\{ (r/\alpha^2)[I_0(\alpha r/2)K_1(\alpha r/2)-
I_1(\alpha r/2)K_0(\alpha r/2)]$$
$$+ (r^2/2\alpha)[I_0(\alpha r/2)K_0(\alpha r/2)+
I_1(\alpha r/2)K_1(\alpha r/2)] \}
\eqno(20)$$

To obtain test particle rotational velocities  we need only differentiate
Eqs. (16) and (20) with respect to $r$. This is readily achieved via
repeated use of the recurrence relations of Eqs. (19) which form a closed
set under differentiation so that higher modified Bessel functions such as
$I_2(\alpha r/2)$ and $K_2(\alpha r/2)$ are not encountered; and the procedure
is
found to yield
$$rV^{\prime}(r)=
(N\beta c^2\alpha^3 r^2/2)[I_0(\alpha r/2)K_0(\alpha r/2)-
I_1(\alpha r/2)K_1(\alpha r/2)]$$
$$+(N\gamma c^2r^2\alpha/2)I_1(\alpha r/2)K_1(\alpha r/2)
\eqno(21)$$
Using the asymptotic properties of the modified Bessel functions we find
that at distances much larger than the scale length $R_0$ Eq. (21) yields
$$rV^{\prime}(r) \rightarrow {N\beta c^2 \over  r}+
{N\gamma c^2r \over 2} -{3N\gamma c^2R_0^2\over 4 r} \eqno(22)$$
as would be expected. We recognize the asymptotic Newtonian term to be just
$N\beta c^2/r$ where $N$ is the number of stars in the disk. The quantity
$N\beta c^2$ is usually identified as $MG$ with $M$ being taken to be the
mass of the disk. For normalization purposes it is convenient to use this
coefficient to define the velocity $v_0=c(N\beta/R_0)^{1/2}$, the velocity
that a test particle would have if orbiting a Newtonian point galaxy with the
same total mass at a distance of one scale length. In terms of the convenient
dimensionless parameter $\eta=\gamma R_0^2/\beta$ Eq. (21) then yields for the
rotational velocity $v(r)$ of a circular orbit in the plane of a thin
exponential
disk the exact expression
$$v^2(r)/v_0^2=(r^2\alpha^2/2)[I_0(\alpha r/2)K_0(\alpha r/2)+
(\eta-1)I_1(\alpha r/2)K_1(\alpha r/2)] \eqno(23)$$
an expression which is surprisingly compact. For thin disks then all departures
from the
standard Freeman result are thus embodied in the one parameter $\eta$ in the
simple manner
indicated.

Beyond making actual applications to galaxies, a further advantage of
having an exact solution in a particular case is that it can be used to test
a direct numerical evaluation of the galactic potential (which involves
integrable infinities) by also running the program for a model exponential
disk. Also, it is possible to perform the calculation analytically in
various other specific cases. For a thin axisymmetric disk with a Gaussian
surface matter distribution $\Sigma(R)=\Sigma_0$exp$(-\alpha^2R^2)$ and
$N=\pi\Sigma_0/\alpha^2$ stars
(this being a possible model for the sometimes steeper central region
of a galaxy in cases where there may be no spherical bulge) we find for the
complete rotational velocity the expression
$$rV^{\prime}(r)=\pi^{1/2}N\beta c^2\alpha^3 r^2
[I_0(\alpha^2r^2/2)-I_1(\alpha^2r^2/2)]e^{-\alpha^2 r^2/2}$$
$$+(\pi^{1/2}N\gamma c^2\alpha r^2/4) [I_0(\alpha^2r^2/2)+
I_1(\alpha^2r^2/2)]e^{-\alpha^2 r^2/2} \eqno(24)$$
\noindent
Similarly, for a spherically symmetric matter distribution (such as the central
bulge region of a galaxy) with radial matter density $\sigma(r)$
and $N=4\pi \int  dr^{\prime}r^{\prime 2}\sigma(r^{\prime})$ stars
a straightforward calculation yields
$$rV^{\prime}(r)={4\pi\beta c^2\over r}\int_0^r dr^{\prime}\sigma(r^{\prime})
r^{\prime 2}$$
$$+{2\pi\gamma c^2\over 3r}\int_0^r dr^{\prime}\sigma(r^{\prime})
(3r^2r^{\prime 2}-r^{\prime 4})
+{4\pi\gamma c^2r^2\over 3}\int_r^{\infty} dr^{\prime}\sigma(r^{\prime})
r^{\prime } \eqno(25)$$
which can readily be integrated once a particular $\sigma(r)$ is specified.
For spherical systems $\sigma(r)$ is not actually measured directly. Rather, it
is
extracted from the surface matter distribution $I(R)$ via an Abel
transform
$$\sigma(r)=-{1 \over \pi} \int _r^{\infty} dR{ I^{\prime}(R)
\over (R^2-r^2)^{1/2}}~~~~,~~~~
I(R)=2\int _R^{\infty} dr {\sigma(r) r \over (r^2-R^2)^{1/2}} \eqno(26)$$
thus leading to double integrals in Eq. (25). However, reduction of these
integrals to one-dimensional integrals over $I(R)$ is possible since we can
rewrite Eq. (25) as
$$rV^{\prime}(r)=-2\beta c^2S(r) + {2\beta c^2\over
r}\int_0^rdr^{\prime}S(r^{\prime})
+\gamma  c^2r
\int_0^rdr^{\prime}S(r^{\prime})
-{\gamma c^2\over r}\int_0^rdr^{\prime }r^{\prime 2}S(r^{\prime})\eqno(27)$$
where the strip brightness $S(x)$ obeys
$$S(x)=2\int_x^{\infty}dR {RI(R) \over (R^2-x^2)^{1/2}}~~~~,
{}~~~~\sigma(x)=-{S^{\prime}(x) \over 2\pi x}\eqno(28)$$
\noindent
Noting that
$${d \over dr}\left\{ 2\int _r^{\infty}dRRI(R) arcsin \left({r \over R} \right)
 \right\}
=-\pi rI(r)+S(r)~~~~,$$
$${d \over dr}\left\{ 2\int _r^{\infty}dRRI(R) \left[ R^2 arcsin \left({r \over
R} \right) -r(R^2-r^2)^{1/2} \right] \right\}
=-\pi r^3I(r)+2r^2S(r) \eqno(29)$$
enables us to conveniently reexpress Eq. (25) directly in terms of $I(R)$, viz.
$$rV^{\prime}(r)=
{4 \beta c^2\over r}\int_r^{\infty}dRRI(R)
\left[ arcsin \left({r \over R} \right) -{r \over (R^2-r^2)^{1/2}}\right] +$$
$${2 \pi \beta c^2\over r} \int_0^rdRRI(R) +
{\gamma c^2 \pi \over 2r} \int_0^rdRRI(R)(2r^2-R^2)$$
$$+{\gamma c^2  \over r}\int_r^{\infty}dRRI(R)
\left[ (2r^2-R^2)arcsin \left({r \over R} \right) +r(R^2-r^2)^{1/2}\right]
\eqno(30)$$
with the Newtonian contribution having previously been noted by Kent (1986).

Beyond the exact expressions obtained above there is one other case of
practical interest
namely that of non-thin but separable disks, a case which can also be greatly
simplified
by our formalism. For such separable disks we set
$\rho(R,z^{\prime})=\Sigma(R)f(z^{\prime})$ where the usually
symmetric thickness function $f(z^{\prime})=f(-z^{\prime})$ is normalized
according to
$$\int_{-\infty}^\infty dz^{\prime}f(z^{\prime})
=2\int_0^\infty dz^{\prime}f(z^{\prime})=1 \eqno(31)$$
\noindent
Recalling that
$$e^{-k\vert z -  z^{\prime}\vert}=\theta(z -  z^{\prime})e^{-k(z -
z^{\prime})}+
\theta(z^{\prime}-z)e^{+k(z -  z^{\prime})} \eqno(32)$$
\noindent
we find that Eqs. (9) and (11) then yield for points with $z=0$
$$V_{\beta}(r)=-4\pi\beta c^2\int_{0}^{\infty} dk\int_{0}^{\infty}dR
\int_0^{\infty}dz^{\prime}R \Sigma(R)f(z^{\prime})
J_0(kr)J_0(kR) e^{-kz^{\prime}}\eqno(33)$$
and
$$V_{\gamma}(r)=2\pi\gamma c^2\int_{0}^{\infty} dk\int_{0}^{\infty}dR
\int_0^{\infty}dz^{\prime}R \Sigma(R)f(z^{\prime})$$
$$\times~[(r^2+R^2+z^{\prime 2})J_0(kr)J_0(kR)
-2rR J_1(kr)J_1(kR)]e^{-kz^{\prime}} \eqno(34)$$
in the separable case. Further simplification is possible if the radial
dependence
is again exponential (viz. $\Sigma(R)=\Sigma_0$exp$(-\alpha R))$ and yields,
following
some algebra involving the use of the recurrence relation
$J_1^{\prime}(z)=J_0(z)-J_1(z)/z$, the expressions
$$rV_{\beta}^{\prime}(r)=2N\beta c^2\alpha^3 r \int_{0}^{\infty}dk
\int_0^{\infty}dz^{\prime}{ f(z^{\prime})e^{-kz^{\prime}}kJ_1(kr)
\over (\alpha^2+k^2)^{3/2}} \eqno(35)$$
and
$$rV_{\gamma}^{\prime}(r)=N \gamma c^2\alpha^3 r \int_{0}^{\infty}dk
\int_0^{\infty}dz^{\prime} f(z^{\prime})e^{-kz^{\prime}}$$
$$\times~ \left( -{4rJ_0(kr) \over (\alpha^2+k^2)^{3/2}}
+{6\alpha^2rJ_0(kr) \over (\alpha^2+k^2)^{5/2}}
-{(r^2+z^{\prime 2})kJ_1(kr) \over (\alpha^2+k^2)^{3/2}}
+{9kJ_1(kr) \over (\alpha^2+k^2)^{5/2}}
-{15\alpha^2kJ_1(kr) \over (\alpha^2+k^2)^{7/2}} \right)
\eqno(36)$$

As regards actual specific forms for $f(z^{\prime})$, two particular ones have
been
identified via the surface photometry of edge on galaxies, one by van der Kruit
and
Searle (1981), and the other by Barnaby and Thronson (1992). Respectively they
are
$$f(z^{\prime})=sech^2(z^{\prime}/z_0)/2z_0 \eqno(37)$$
and
$$f(z^{\prime})=sech(z^{\prime}/z_0)/\pi z_0 \eqno(38)$$
each with appropriate scale height $z_0$. (A recent faint starlight search by
Sackett et. al. 1994
suggests that for NGC 5907 the above previously detected exponential drop
gradually
softens into a power-law behavior at larger scale heights. The effect of this
tail on fitting
will be negligible if it possesses the same mass to light ratio as all the
other matter that had
previously been detected in the galaxy, and so we shall not consider it further
here). We note that
both of the thickness functions of Eqs. (37) and (38)
are falling off very rapidly in the $z^{\prime}$ direction just like the
Bessel function expansion itself of Eq. (5). Consequently,
Eqs. (33) - (36) will now have very good convergence properties.
The thickness function of Eq. (37) is found to lead to rotational velocities of
the form
$$rV^{\prime}_{\beta}(r)=
(N\beta c^2\alpha^3 r^2/2)[I_0(\alpha r/2)K_0(\alpha r/2)-
I_1(\alpha r/2)K_1(\alpha r/2)]$$
$$-N \beta c^2\alpha^3 r\int_0^\infty dk {k^2J_1(kr)z_0\beta(1+kz_0/2) \over
(\alpha^2+k^2)^{3/2}} \eqno (39)$$
where
$$\beta(x)=\int_0^1{t^{x-1} \over (1+t)} \eqno(40) $$
and
$$rV_{\gamma}^{\prime}(r)=N\gamma c^2\alpha^3 r \int_{0}^{\infty}dk
(1-kz_0\beta(1+kz_0/2))$$
$$\times~ \left( -{2rJ_0(kr) \over (\alpha^2+k^2)^{3/2}}
+{3\alpha^2rJ_0(kr) \over (\alpha^2+k^2)^{5/2}}
-{r^2kJ_1(kr) \over 2(\alpha^2+k^2)^{3/2}}
+{9kJ_1(kr) \over 2(\alpha^2+k^2)^{5/2}}
-{15\alpha^2kJ_1(kr) \over 2(\alpha^2+k^2)^{7/2}} \right)$$
$$+N\gamma c^2\alpha^3 r \int_{0}^{\infty}dk{kJ_1(kr)
\over 2(\alpha^2+k^2)^{3/2}}
{d^2 \over dk^2}\left( kz_0\beta(1+{kz_0 \over 2})
\right) \eqno(41)$$
Similarly, the thickness function of Eq. (38) leads to
$$rV^{\prime}_{\beta}(r)={2N\beta c^2\alpha^3 r \over \pi}
\int_0^\infty dk {kJ_1(kr)\beta(1/2+kz_0/2) \over (\alpha^2+k^2)^{3/2}}
\eqno (42)$$
and
$$rV_{\gamma}^{\prime}(r)={N\gamma c^2\alpha^3 r \over \pi} \int_{0}^{\infty}dk
\beta(1/2+kz_0/2)$$
$$\times~ \left( -{4rJ_0(kr) \over (\alpha^2+k^2)^{3/2}}
+{6\alpha^2rJ_0(kr) \over (\alpha^2+k^2)^{5/2}}
-{r^2kJ_1(kr) \over (\alpha^2+k^2)^{3/2}}
+{9kJ_1(kr) \over (\alpha^2+k^2)^{5/2}}
-{15\alpha^2kJ_1(kr) \over (\alpha^2+k^2)^{7/2}} \right)$$
$$-{N\gamma c^2\alpha^3 r \over \pi} \int_{0}^{\infty}dk{kJ_1(kr)
\over (\alpha^2+k^2)^{3/2}}
{d^2 \over dk^2}\left(\beta({1+kz_0 \over 2}) \right)
\eqno(43)$$
\noindent
The great utility of these expressions is that all of the functions of
$\beta(x)$
and their derivatives which appear in Eqs. (39) - (43) converge very rapidly to
their asymptotic values as their arguments increase. Consequently the $k$
integrations in Eqs. (39) - (43) converge very rapidly numerically while
encountering no singularities at all.

As a practical matter, the observed scale heights $z_0$ are usually much
smaller than
any observed scale lengths $R_0$. Consequently the thickness corrections of
Eqs.
(39) - (43) usually only modify the thin disk formula of Eq. (21) in the
central galactic
region, and thus have essentially no effect on the linear potential
contribution.
For the Newtonian term
the corrections of Eqs. (39) and (42) to the Freeman formula tend to reduce the
overall
Newtonian contribution (c.f. the sign of the integral term in Eq. (39) and
Casertano 1983)
and serve to ensure that the inner rotation curves of Fig. (1)
are well described (see Mannheim 1993b) by the luminous Newtonian contribution,
to thus
clear the way to explore the effect of the linear term on the outer region of
the
rotation curve, a region where its presence is significant and where the thin
disk formula
of Eqs. (21) and (23) provides a very good approximation to the dynamics.
\medskip
\noindent
{\bf (3) Some General Systematics of Galactic Rotation Curve Fitting}
\medskip
In order to understand the general features of the rotation curves of Fig. (1)
it is
instructive to consider the generic implications of the thin disk formula of
Eq. (23), a two
parameter formula with $v_0$ fixing the overall normalization and $\eta$ the
relative
contributions of the Newtonian and linear pieces. Moreover, if this (per
galaxy) overall
normalization is fixed by the peak in the rise of the inner rotation curve (the
so called maximum
disk fit in which the Newtonian disk contribution gets to be as large as it
possibly can be),
then essentially the entire shape of the rest of the curve is fixed by just the
one parameter
(per galaxy) $\eta$. As regards this maximum disk contribution, we note that
the Newtonian
term in Eq. (23) peaks at 2.15$R_0$ with $v^2/v_0^2$ receiving a Newtonian
contribution of
0.387. This Newtonian contribution comes down to half of this value (i.e.
0.194) at 6.03$R_0$.
Since the linear contribution is essentially negligible at 2.15$R_0$
(especially after we take
the square root to get the velocity itself), if we choose the linear
contribution at 6.03$R_0$
to be equal to the Newtonian contribution at that same distance (i.e. if
numerically we set
$\eta$ equal to a critical value of 0.067, to thus fix our one free parameter
(per galaxy)
once and for all), we will then have essentially achieved flatness over the
entire 2 to 6 scale
length region. Now at 6 scale lengths both the Newtonian and linear terms are
quickly approaching
the asymptotic values exhibited in Eq. (22), and at close to 12 scale lengths
(precisely at
11.62$R_0$) the linear term contribution to $v^2/v^2_0$ is just 0.387, the
original maximum disk
value at 2.15$R_0$. Consequently, between 6 and 12 scale lengths the rotation
curve will again show
little deviation from flatness without any further adjustment of parameters at
all. However since
the Newtonian contribution at 12 scale lengths is slightly bigger than the
linear contribution at 2
scale lengths, the net outcome is that by 12 scale lengths the rotation curve
is actually beginning
to show a slight rise, with flatness only being achieved out to about 10 scale
lengths.
Thus in general we see that by varying just one parameter we can naturally
achieve flatness over the entire 2 to 10 scale length region, this intriguingly
being about as
large a range of scale lengths as has up till now been observed in any rotation
curve.
In order to see just how flat a rotation curve it is in principle possible to
obtain, we have varied
$\eta$ as a free parameter. Our most favored generic case is then obtained when
$\eta$ takes the value 0.069 (i.e. essentially the critical value),
with the resulting generic rotational velocity curve being
plotted in Fig. (2). Over the range from 3 to 10 scale lengths the ratio
$v(r)/v_0$ is found
to take the values (0.666, 0.648, 0.632, 0.626, 0.628, 0.637, 0.651, 0.667) in
unit step
increases. Thus it has a spread of only $\pm 3 \%$ about a central value of
0.647 in this region.
Additionally, we find that even at 15 scale lengths the ratio $v(r)/v_0$ has
still only
increased to 0.763, a 14$\%$ increase over its value at 10 scale lengths. In
the upper
diagram in Fig. (2) we have plotted the generic $\eta=0.069$ rotation curve out
to 10 scale
lengths to show just how flat it can be. In the lower diagram in Fig. (2) we
have shown the
continuation out to 15 scale lengths where the ultimate asymptotic rise is
becoming apparent.
We have deliberately juxtaposed the two diagrams in Fig. (2) since the flatness
out to 10
scale lengths is usually taken as being indicative of asymptotic flatness as
well, with such
ultimate flatness being characteristic (and even a primary motivation)
of both isothermal gas sphere dark matter models and the MOND alternative
(Milgrom 1983a, b, c). The possibility that flatness is only
an intermediate and not an asymptotic phenomenon is one of the most unusual and
distinctive
features of the conformal gravity theory. (Of course it is always possible to
build dark
matter models with non flat asymptotic properties (see e.g. van Albada et al.
1985) since
the dark matter theory is currently so unconstrained. However, our point here
is that the
conformal theory is the first theory in which rotation curves are actually
required to
ultimately rise, even being predicted to do so in advance of any data). As
regards other
possible values for $\eta$, if $\eta$ exceeds the critical value of 0.067, then
the curve
will be flat for fewer scale lengths with the rise setting in earlier (given
the large
value for $R_0$, and hence  $\eta$, that the galaxy UGC 2885 happens to possess
(see below),
we note in passing that a study of its outer region might thus provide a good
opportunity to
detect such a possible rise); while if $\eta$ is less than the critical value,
the curve will
drop perceptibly before coming back to its maximum disk value at a greater
distance.

As regards the generic critical value for $\eta$, we note that for a typical
galaxy with a mass of
10$^{11}$ solar masses and a 3 kpc scale length, the required value for the
galactic
$\gamma_{gal}~(=N\gamma_{star}$ where $\gamma_{star}$ is the typical $\gamma$
used in the stellar potential $V(r)$ of Eq. (3)) then turns out to be of order
10$^{-29}$/cm, which, intriguingly, is of order the inverse Hubble radius.
Moreover, this
characteristic value is in fact numerically attained in the fits of Fig. (1)
for the stellar disk contribution in all of our four
chosen galaxies  (viz. $\gamma$(154)=2.5$\times$10$^{-30}$/cm,
$\gamma$(3198)=3.5$\times$10$^{-30}$/cm,
$\gamma$(2903)=7.6$\times$10$^{-30}$/cm,
$\gamma$(5907)=5.7$\times$10$^{-30}$/cm). (While this same cosmological value
is also found for
DDO 154, in some other aspects (such as possessing a rotation curve which has
no observed flat
region at all) the fitting to this dwarf irregular is found
to be anomalous (see Mannheim 1993b) presumably because the galaxy is gas
rather than star
dominated. Hence we shall only regard the three other galaxies, all regular
spirals, as typical
for the purposes of our discussion here). Thus not only is $\gamma_{gal}$
making the
observed representative curves flat in the observed region, and not only is it
doing so with an effectively universal value, it is doing so with a value which
is already
known to be of astrophysical significance; thereby suggesting that
$\gamma_{gal}$ may be
of cosmological origin, perhaps being
related to the scale at which galaxies fluctuate out of the cosmological
background.

Additionally, we note that this apparent universality for $\gamma_{gal}$ has
implications for
the status of the Tully-Fisher relation in our theory. Specifically, the
average velocity $v_{ave}$
of the flat part of the critical generic curve (but only of the flat part since
the curve
must ultimately rise in our theory) is equal to the maximum
disk value since the curve is flat in that region. Thus we can set (using
$N=2\pi\Sigma_0 R_0^2$
and letting $L$ denote the galactic luminosity)
$$v_{ave}^4=\left( {0.387N\beta c^2\over R_0} \right)^2
=0.300\pi\Sigma_0\beta^2 c^4 L\left( {N \over L} \right)
\eqno(44)$$
\noindent
At the critical value for $\eta$
(the fits yield $\eta$(3198)=0.044, $\eta$(2903)=0.038, $\eta$(5907)=0.057) we
also can set
$$\gamma_{gal} c^2={ 0.067N\beta c^2\over R_0^2}=0.134\pi\Sigma_0\beta
c^2\eqno(45)$$
so that Eq. (44) may be rewritten as
$$v_{ave}^4=2.239 \gamma_{gal}\beta c^4 L\left( {N \over L} \right) \eqno(46)$$
\noindent
If we assume that all galaxies possess the same universal value for the disk
mass to light ratio
(our fits yield $M/L$(3198)=4.2, $M/L$(2903)=3.5, $M/L$(5907)=6.1 in units of
$M_{\odot}/L_{B\odot}$), we then see that given a universal $\gamma_{gal}$ and
a universal $\eta$,
Eq. (46) then yields
noneother than the Tully-Fisher velocity-luminosity relation. (Observationally
the Tully-Fisher
relation is not thought to hold for the stellar component of the dwarf
irregular DDO 154, as may be
anticipated since DDO 154 is phenomenologically found to have an anomalously
small $M/L$
ratio ($M/L$(154) takes the value 1.4 in our fits and is essentially zero in
the dark matter and
MOND fits of Begeman, Broeils, and Sanders 1991) - since the above discussion
does not include any
non-stellar component it is anyway not applicable to gas dominated galaxies).
Additionally,
according to Eq. (45) the universality of $\gamma_{gal}$ also entails the
universality of
$\Sigma_0$, the central surface brightness, an as yet unexplained
phenomenological feature first
identified for spirals by Freeman (1970). (In turn the universality of
$\Sigma_0$ entails a
mass - radius squared relation for galaxies). The (near) universality of
$\gamma_{gal}$ and of
$\eta$ (the near universality found for $\eta$ is accounted for by the
phenomenological fact that
the scale lengths $R_0$ of the three spiral galaxies in our sample are all
quite close to each
other) thus correlates in one fell swoop the observed flatness of rotation
curves, the universality
of $\Sigma_0$, and the Tully-Fisher relation, and does so in a theory in which
rotation curves must
eventually rise. As such, the above given discussion provides
a generalization to axially symmetric systems of an earlier discussion (Kazanas
1991, Kazanas and
Mannheim 1991b) based on the simplification of using Eq. (2) itself as the
galactic metric. As
we now see, the ideas developed in those two papers carry over to the present
more detailed
treatment. (In passing we should point out the mass - radius squared relation
which was also
identified in those two previous papers was actually found to have
phenomenological validity on many
other astrophysical scales as well, something which still awaits an
explanation).

It is important to note that the above discussion does not constitute a
complete first principles
derivation of the Tully-Fisher relation. Rather, even while doing so in an
unforced and natural
way, the discussion nonetheless takes advantage of the phenomenological facts
that $\Sigma_0$ and
$R_0$ are each quite close to universal without explaining why this is so.
(Universal $\Sigma_0$
entails Tully-Fisher for the maximum disk peak velocity in galaxies which have
no significant
inner region bulge, while universal $\eta$ then extends the Tully-Fisher
relation to the average
rotation velocity to the extent that the curve is in fact flat.) However, since
Eq. (45) does
correlate $\Sigma_0$ with $\gamma_{gal}$ and since $\gamma_{gal}$ is
numerically of cosmological
significance, our analysis does suggest that the theoretical establishing of a
cosmological origin
for $\gamma_{gal}$ in which it would emerge as being of order the inverse
Hubble radius (presumably
in a theory of galaxy formation via cosmological inhomogeneities) would then
lead naturally to
intermediate region flatness, to the Tully-Fisher relation, and to the
universality of $\Sigma_0$.
As regards the Tully-Fisher relation in general, we note that it actually has
two aspects to it, one
being the obvious fact that it is the fourth power (as opposed to any other
possible power) of the
velocity which is universally related to the luminosity, and the other being
the much deeper fact
that there is actually a universal correlation at all, i.e. that the velocity
(which is fixed by the
total gravitational potential) is correlated with the luminosity (which is
fixed by the visible
matter alone) rather than being correlated with any possible non-luminosity.
Thus the very
fact that there is a Tully-Fisher relation at all thus quite strongly suggests
that galaxies should
therefore be predominantly luminous, and that it is the rule for calculating
the potential which
hence needs to be changed. Moreover, our ability to obtain Eqs. (44-46) and the
fits of Fig. (1) in
our theory follows precisely because the linear potential is integrated over
the same luminous
matter distribution as the Newtonian potential, to thus automatically normalize
both the
contributions (and hence the total velocity) to one and the same luminosity in
a completely natural
manner.

While we have categorized our fits as having two parameters per galaxy, the
actual situation is
slightly more constrained. Specifically, we note that the Newtonian and linear
contributions are
both proportional to $N$ according to Eq. (22). Thus if there existed universal
average stellar
parameters $\beta$ and $\gamma$ to serve as input for Eqs. (4) and (11), $\eta$
would then be fixed
by the scale length $R_0$ of each galaxy, resulting in one parameter ($N$) per
galaxy fits.
Ordinarily, one thinks of $\beta$ as being the Schwarzschild radius of the Sun,
and then in the fits
the numerical value of the mass to light ratio of the galaxy is allowed to vary
freely in the
fitting, with disk $M/L$ ratios then being found which are actually remarkably
close to each other
(without such closeness there would be severe violations of the Tully-Fisher
relation because of the
$N/L=M/LM_{\odot}$ factor in Eqs. (44) and (46)). However, in reality each
galaxy comes with its own
particular mix of stars, both in overall population and, even more
significantly, in the spatial
distribution of the mix. Now, of course ideally we should integrate Eq. (4)
over the true stellar
distribution allowing $\beta$ to vary with position according to where the
light and heavy stars
(stars whose luminosities do not simply scale linearly with their masses) are
physically
located within the stellar disk. Instead we use an average $\beta$ (which
incidentally enables us to
derive exact formulas such as Eq. (9)). However, two galaxies with the
identical morphological mix
of stars but with different spatial distributions of those stars should each be
approximated by a
different average $\beta$, since the Newtonian potential weights different
distances unequally.
Since we do not give two galaxies of this type different average $\beta$
parameters to begin with,
we can then compensate later by giving them different mass to light ratios
(even though for this
particular example we gave them the same morphological mix). Hence we extract
out a quantity
$N\beta_{\odot}$ from the data which simulates $N_{ave}\beta_{ave}$ where
$N_{ave}$ is the true
average number of stars in the galaxy. Because of the difference between these
two ways of
defining the number of stars in a galaxy, it is not clear whether the currently
quoted mass to
light ratios as found in the fits (in essentially all theories of rotation
curve systematics)
are merely reflecting this difference or whether they are exploiting this
uncertainty to
come up with possibly unwarrantable mass to light ratios. Thus a first
principles determination of
actual values or of a range of allowed values of galactic disk mass to light
ratios prior to fitting
would be extremely desirable, with (as we shall see below) the recent round of
microlensing
observations actually taking a first step in this direction.

A precisely similar situation also obtains for the $\gamma$ dependent terms.
Again we use an
average stellar $\gamma$ and compensate for its possible average variation from
galaxy to galaxy
by allowing the galactic disk gamma to light ratio
($\gamma_{gal}/L=N\gamma/L$), and hence the
galactic $\eta$, to vary phenomenologically (i.e. we use $N\gamma$ to simulate
$N_{ave}\gamma_{ave}$ where $N$ is determined once and for all by normalizing
the data to
$N\beta_{\odot}$). The fits to our representative galaxies are found to yield
$N\gamma/L_B$(3198)=3.9,
$N\gamma/L_B$(2903)=5.1, $N\gamma/L_B$(5907)=3.2 (in units of
$10^{-40}$/cm/$L_{B\odot}$),
values which again are remarkably close to each other and which are of a par
with the mass to light
ratios $M/L$(3198)=4.2, $M/L$(2903)=3.5, $M/L$(5907)=6.1 found for the same
galaxies. We would not
expect the $M/\gamma_{gal}$ ratio to be the same for the entire sample, simply
because even if
the stellar $\beta$ and $\gamma$ parameters were to change by the same
proportion in going from one
morphological type of star to another (a reasonable enough expectation),
nonetheless, as the
galactic spatial distributions change, the inferred average stellar $\beta$ and
$\gamma$ parameters
would then change in essentially unrelated ways, since the Newtonian and linear
potentials weight
the differing spatial regions of the galaxy quite differently to thus yield
different average
values. Nonetheless, it is intriguing to find that the variation in the average
$\beta$ and $\gamma$
shows such mild dependence on specific galaxy within our sample; $N$ and
$N_{ave}$ thus appear to
be very close, with this small variation absorbing the variation in scale
length $R_0$ used to
determine the various $\eta$. To within this (mild) variation, our fits are
thus effectively one
parameter per galaxy fits. (In its pure form the MOND explanation of the
systematics of galactic
rotation curves is also a one parameter per galaxy theory. However, in its
successful practical
applications (Begeman, Broeils, and Sanders 1991), it is generally found
necessary to introduce at
least one more fitting parameter per galaxy, such as by allowing a (generally
quite
mild) variation in the fundamental acceleration parameter $a_0$ over the
galactic sample.
Phenomenologically then MOND would thus appear to be on a par with our linear
potential theory).
For our linear potential theory we note that given the apparent uniformity of
the average stellar
$\gamma/\beta$ ratio, we see that we really have to normalize $N$ to the
maximum disk mass and that
we are really not free to vary the normalizations of the Newtonian and linear
pieces separately,
since they both are proportional to $N$. Specifically, if we make the
Newtonian piece too small we would have to arbitrarily increase the linear
contribution, something
we are not able to do in a consistent manner. Thus the Newtonian contribution
in our fit cannot be
too small. Similarly, it can never be allowed to be too large (this would give
too high a velocity);
and, hence, the Newtonian contribution in our theory is bounded both above and
below, and
essentially forced to the maximum disk mass; and thus our theory is reduced to
almost parameter
free fitting. Since dark matter fits can generally adjust the relative
strengths of the luminous
and dark matter pieces at will, they are not so constrained, and often yield
much smaller luminous
Newtonian contributions, and thus large amounts of dark matter, particularly in
fits to dwarf
galaxies. Thus a first principles determination of galactic disk mass to light
ratios might enable
one to discriminate between rival theories. (As we discuss below this is
precisely beginning to
happen via the microlensing observations, with the initial data apparently even
supporting the
conformal gravity maximum disk, minimal halo scenario.)

In order to compare our work with that of other approaches it is useful
to clarify the significance of the term `flat rotation curve'. In the
literature it is generally thought that rotation curves will be flat
asymptotically (though of course the most significant aspect of the data
is the fact that the curves
deviate from the luminous Newtonian prediction at all, rather than in what
particular way);
and of course since our model predicts that velocities will
eventually grow as $r^{1/2}$, the initial expectation is that our model
is immediately ruled out. However, the rotation curve fits that have so far
been made are
not in fact asymptotic ones. Firstly, the $HII$ optical studies pioneered by
Rubin
and coworkers (Rubin et al. 1970, 1978, 1980, 1982, 1985), even while they were
indeed yielding
flat rotation curves, were restricted to the somewhat closer in optical disk
region since the $HII$ regions are only to be found in the vicinity of hot
stars which ionize those
regions. And eventually, after a concentrated effort to carefully measure the
surface brightnesses
of such galaxies, it gradually became apparent (see e.g. Kaljnas 1983 and Kent
1986) that the $HII$
curves could be described by a standard luminous Newtonian prediction after
all; even in fact for galaxies such as UGC 2885 for which the data go out to as
much as 80 kpc, a
distance which turns out to only be of order 4 scale lengths ($R_0$=22 kpc for
UGC 2885, an
atypically high value - this galaxy is just very big.)  Thus, not only are the
optical studies
limited (by their very nature in fact) to the optical disk region where there
is some detectable
surface brightness, but, albeit by coincidence, it turns out that they are also
limited to the
region where an extended Newtonian source is actually yielding flat rotation
curves to a rather
good degree. Thus this inner region flatness has nothing at all to do with any
possible asymptotic
flatness, though it will enable flatness to set in as early as 2 or 3 scale
lengths in fits
to any data which do go out to many more scale lengths.

While the $HII$ data do not show any substantive non-canonical behavior,
nonetheless,
the pioneering work of Rubin and coworkers brought the whole issue of galactic
rotation
curves into prominence and stimulated a great deal of study in the field. Now
it turns
out that neutral hydrogen gas is distributed in galaxies out to much farther
distances than
the stars, thus making the $HI$ studies ideal probes of the outer reaches of
the rotation
curves and of the luminous Newtonian prediction. (That $HI$ studies might lead
to a conflict with
the luminous Newtonian prediction was noted very early by Freeman 1970 from an
analysis of
NGC 300 and M33, by Roberts and Whitehurst 1975 from an analysis of M31, and by
Bosma 1978, 1981
who made the first large 21 cm line survey of spiral galaxies).
Thus with the $HI$ studies (there are now about 30 well studied cases) it
became clear that there really was a problem with the interpretation of
galactic rotation
curve data, which the community immediately sought to explain by the
introduction of
galactic dark matter since the Newton-Einstein theory was
presumed to be beyond question. (So much so in fact that Ostriker and Peebles
1973 had already
introduced a spherical dark matter halo to stabilize otherwise unstable
Newtonian disk galaxies).
Fits to the $HI$ data have been obtained using dark matter
(Kent 1987 provides a very complete analysis),
and while the fits are certainly phenomenologically acceptable, they
nonetheless possess
certain shortcomings. Far and away their most serious shortcoming is their ad
hoc nature,
with any found Newtonian shortfall then being retroactively fitted by an
appropriately
chosen dark matter distribution. In this sense dark matter is not a predictive
theory
at all but only a parametrization of the difference between observation and the
luminous
Newtonian expectation. As to possible dark matter distributions, no specific
distribution,
or explicit set of numerical parameters for a distribution, has
convincingly been derived from first principles as a consequence, say, of
galactic
dynamics or formation theory (for a recent critical review see Sanders 1990).
(The general community would not appear to regard any specific derivation as
being all that
convincing since no distribution has been heralded as being so theoretically
secure
that any failure of the data to conform to it would obligate the community to
have to abandon
the Newton-Einstein theory).
Amongst the candidate dark matter distributions which have been considered in
the literature
the most popular is the distribution associated with a modified isothermal gas
sphere (a two parameter spherical matter volume density distribution
$\rho(r)=\rho_0/(r^2+r_0^2)$
with an overall scale $\rho_0$ and an arbitrarily introduced non-zero core
radius $r_0$
which would cause dark matter to predominate in the outer rather than the inner
region -
even though a true isothermal sphere would have zero core radius). The appeal
of the
isothermal gas sphere is that it leads to an asymptotically logarithmic
galactic
potential and hence to asymptotically flat rotation curves, i.e. it is
motivated by no less than the
very data that it is trying to explain. However, careful analysis of the
explicit dark
matter fits is instructive. Recalling that the inner region (around, say,
2$R_0$ for
definitiveness) is already flat for Newtonian reasons, the dark matter
parameters are then
adjusted so as to join on to this Newtonian piece (hence the ad hoc core radius
$r_0$) to give a
continuously flat curve in the observed region, rather than one which either
rises or falls to its
presupposed eventual asymptotic flat value. This matching of the luminous and
dark matter pieces is
for the moment completely fortuitous (van Albada and Sancisi 1986 have even
referred to it as a
conspiracy) and not yet explained by galactic dynamics, even though it is only
by such
(assumed universal) matching galaxy by galaxy of the inner region Newtonian
peak velocity to
the presupposed constant asymptotic velocity that the dark matter models can
achieve compatibility
with the universal Tully-Fisher relation, with this treatment of the
Tully-Fisher relation
standing in sharp contrast to that provided by the conformal gravity theory
which was discussed
above. What is done in the dark matter fits is actually even a double
conspiracy. Not only are the
outer (10$R_0$) and the inner (2$R_0$) regions given the same velocity (by
adjusting $\rho_0$), the
intermediate (6$R_0$) region is adjusted through the core radius $r_0$ to
ensure that the curve does
not fall and then rise again in that region. Hence flatness in the $r_0$
dominated region has almost
nothing at all to do with the presumed asymptotically flat isothermal gas
sphere contribution. Even
worse, in the actual fits the dark matter contributions are found to actually
still be rising at the
largest observed (10$R_0$) distances, and thus not yet taking on their
asymptotic values at all.
Hence the curves are made flat not by a flat dark matter contribution but
rather by an interplay
between a rising dark matter piece and a falling Newtonian one (an effect which
is completely
mirrored in the linear potential theory fits of Fig. (1)), with the
asymptotically flat expectation not yet actually having even been tested.
(Prospects for
pushing the data out to farther distances are not good because $HI$ surface
densities
typically fall off exponentially fast at the edge of the explored region.) Thus
for the moment,
even though both available $HI$ and $HII$ type data sets are flat in their
respective domains,
each data set is flat for its own coincidental reason, and it would appear to
us that the
region of true galactic
asymptotics has yet to be explored; with the observed flatness of the galactic
rotation curves
(just like the apparent flatness of total proton proton
scattering cross sections over many energy decades before an eventual rise)
perhaps only being
an intermediate rather than an asymptotic phenomenon.

As regards the conformal gravity theory fits, we see that is not in fact
necessary to demand
flatness in the asymptotic region in order to obtain flat rotation curves in
the explored
intermediate region. Thus, unlike dark matter fitting, we do not need to know
the structure of the
data prior to the fitting (when Mannheim and Kazanas first set out to analyze
conformal gravity
they had no idea what the fits might eventually look like at all), or need to
adapt the model to any
presupposed asymptotic flatness. Our linear potential theory is also more
motivated theoretically
than the dark matter models since Eq. (3) arises in a fundamental, fully
covariant, uniquely
specified theory, while, despite all the attention it has been given, the dark
matter spherical halo
remains an ad hoc assumption.  Additionally, the linear potential is able
(Christodoulou 1991) to
stabilize disk galaxies without any need for dark matter, and the conformal
gravity theory
thus appears to be able to reproduce all the desirable aspects of galactic dark
matter without
needing the dark matter itself. Further, the conformal theory possesses one
fewer free parameter
per galaxy compared to dark halo models ($M/L$ and $\gamma$ instead of $M/L$,
$\rho_0$ and $r_0$).
Consequently, according to the usual criteria for evaluating rival theories, as
long as conformal
gravity continues to hold up, it is to be preferred.
\medskip
\noindent
{\bf (4) Implications of Conformal Gravity for Clusters of Galaxies}
\medskip
In discussing the behavior of gravitating systems with a large number of
degrees of freedom such as
a cluster of galaxies, usually only average information such as a mean velocity
dispersion is
available rather than detailed features such as rotation curves which describe
the motions of the
individual constituents of	the system. Consequently, the analysis is a strictly
statistical one, and
generally is based on the use of equations such as the collisionless Boltzmann
equation and the
assumption of virialization (see e.g. Binney and Tremaine 1987 for a
comprehensive review). Since
the standard discussion generally considers the two-body collision dependent
term in the Boltzmann
equation to be a local perturbation on the global mean field set up by the
gravitational field of
the rest of the particles in the cluster (treating each galaxy as a point
particle for simplicity),
we have to reexamine the entire formalism in light of the linear potentials of
Eq. (3) which grow
with distance and which can thus not be thought of as producing localizable
collisions at all.
Moreover, in a theory with rising potentials, we are not free to ignore the
effects due to the
particles in the rest of the Universe, so that even if a system such as a
cluster of galaxies is
geometrically isolated, that does not immediately mean that it is
gravitationally isolated or that
it is bound purely under its own self forces (something that of course would be
the case in a
strictly Newtonian theory where forces do fall with distance). Thus in order to
fully apply our
theory to clusters we not only need to develop an appropriate formalism for
describing the local
gravitational effects purely within a cluster, but in principle we also need to
consider the
coupling of the cluster to the rest of the Universe. To discuss the effect of
such global coupling
requires developing a theory for the growth of inhomogeneities and galaxy
formation, with the
relevant issue for motions within clusters then being not so much the coupling
of each cluster to
the general Hubble flow produced by a homogeneous background distribution of
sources, but rather
its coupling to the deviations from that flow caused by the presence of
inhomogeneities. Since a
theory for the growth of inhomogeneities in conformal gravity has yet to be
developed, we are
currently unable to address this issue explicitly or explore its numerical
consequences for cluster
velocity dispersions (as a first step though we will determine below the region
where the cluster
density falls to the general cosmological background density and use this to
ascertain where it is
that the cluster actually ends). Consequently, for the purposes of this paper,
we concentrate
primarily on the local gravitational problem within a given cluster, and
actually present
a solution to this problem in the linear potential case based on the
Bogoliubov, Born, Green,
Kirkwood, and Yvon (BBGKY) hierarchy (this being a far more general statistical
formalism than the
collisionful Boltzmann equation to which it can actually reduce under specific
assumptions (see
e.g. Huang 1987 and Liboff 1990)). We first present some general features
of the formalism, and then make an explicit application of it to the Coma
cluster.

Before discussing this Liouville equation based statistical analysis, it is
convenient to
first examine some features which follow purely from the equations of motions
of the cluster
particles as they move in some general potential $V(\bar{x})$ according to
$${d^2 \bar{x}\over dt^2} = -{\partial V(\bar{x}) \over \partial \bar{x}}
\eqno(47)$$
(While we shall of course base our discussion on the non-relativistic geodesics
of Eq. (47),
we note in passing that while Eq. (47) correctly describes the coupling of a
massive test
particle to the Newtonian and linear potentials of Eq. (3), it does not
incorporate any effect due
to a direct interaction between the linear potentials of differing particles.
In a more complete
treatment such an effect would be generated via the two-body correction to
one-body geodesic motion,
an effect which for non-relativistic motions would be expected to only be a
small perturbation on
Eq. (47). Since the two-body problem in fourth order gravity is as far from
solution as that in the
standard second order theory, we shall simply ignore any such effects here.)
Given Eq. (47), it then trivially follows that
$${1 \over 2} {d^2  \over dt^2} \left( \bar{x}^2 \right)
=v^2-r{\partial V(\bar{x})\over \partial r}
\eqno(48)$$
\noindent
Thus for a cluster of $N$ particles (with coordinates labeled by $\alpha$
($\alpha=1,...,N$),
and equal mass $m$ for simplicity), the trace of the moment of inertia tensor
of the cluster obeys
$${1 \over 2}{d^2I \over dt^2}={1 \over 2}{d^2 \over dt^2} \left(
m\sum_{\alpha=1}^{N}r_{\alpha}^2
\right)=   m\sum_{\alpha=1}^{N}v_{\alpha}^2-
m\sum_{\alpha=1}^{N}r_{\alpha}{\partial V(\bar{x}_{\alpha}) \over \partial
r_{\alpha}}
\eqno(49)$$
and thus directly yields for a spherically symmetric system with matter
volume density $\sigma(r)$ the familiar
$${1 \over 2}{d^2I \over dt^2}=4\pi m \int_0^{\infty}drr^2\sigma(r)v^2(r)
-4\pi m \int_0^{\infty}drr^2\sigma(r)rV^{\prime}(r)
\eqno(50)$$
\noindent
Since Eq. (50) is based only on spherical symmetry with no commitment as to the
explicit structure
of the potential being needed, it thus immediately holds both in the Newtonian
case and in our
linear potential case as well, with the appropriate $rV^{\prime}(r)$ needed for
Eq. (50) then being
given directly by Eq. (25) which was explicitly derived earlier for the
spherical case. Finally,
should the entire cluster be in a steady state with $\ddot{I}=0$ (this is
actually the key
assumption as we will see below), we then obtain an expression for the
spatially averaged mean
square virial velocity (here and throughout `av' will mean averaged with
respect to the spatial
distribution)
$$N(v^2)_{av}=4\pi \int_0^{\infty}drr^2\sigma(r)rV^{\prime}(r)
\eqno(51)$$
in the spherically symmetric case, which can then readily be evaluated once an
appropriate
$\sigma(r)$ is specified. (Even if $\dot{I}$ is not in
fact time independent, as long as $\dot{I}$ remains bounded,  Eq. (51) would
still be valid provided
it is reinterpreted as a long time average).

For our discussion of clusters below, we note that
Eq. (51) admits of a local generalization. Since $r^2$ is time independent in a
circular orbit, these
orbits actually satisfy Eq. (51) orbit by orbit and not merely in a statistical
sense. Thus both the
set of all circular orbits and the set of all non-circular ones must each
satisfy Eq. (51)
separately (the circular ones individually, but the non-circular ones only
statistically). Thus suppose that some interior region (say from $r=0$ to some
maximum $r=r_m$ with $N(r_m)$ particles) of the cluster has the property that
all of its
non-circular orbits actually stay within the region with little radial
inflow or outflow between this region and the rest of the cluster. In this
interior region then the set of all of the orbits (circular and non-circular
combined) would then satisfy $\ddot{I}(r<r_m)=0$ and yield the local
$$N(r_m)(v^2(r_m))_{av}=
4\pi \int_0^{r_m}drr^2\sigma(r)rV^{\prime}(r)
\eqno(52)$$
for a cluster whose central region is in a steady state. For a Newtonian system
Eq. (52)
is completely self-contained since the total potential then only depends on the
matter
interior
to the point of observation with Eq. (52) only receiving contributions from the
matter interior
to $r_m$. However, as can be seen from Eq. (25), in the linear potential case
the total
potential at a point is also sensitive to the
matter exterior to that point, and thus the rest of the cluster exterior to
$r_m$ (and, in principle,
the rest of the matter in the Universe as well) contributes non-trivially and
potentially even
significantly to the virial velocity within any virialized central region of
radius $r_m$. As we
will see below, the crucial issue in applying the virial to clusters will turn
out to be determining
just how big $r_m$ might be, and in order to see how to address this point we
turn to a statistical
BBGKY analysis.

For a system of $N$ particles moving under the conservative forces associated
with equations of
motion such as Eq. (47), the normalized (to one) 6$N$ dimensional distribution
function
$f^{(N)}(\bar{w}_{\alpha},t)$ ($\bar{w}_{\alpha}= \{\bar{x}_{\alpha},~
\bar{v}_{\alpha}\}$) obeys the Liouville equation
$$ {df^{(N)} \over dt}={\partial f^{(N)} \over \partial t}+\sum_{\alpha=1}^{N}
\left[
\bar{v}_{\alpha} \cdot {\partial f^{(N)} \over \partial \bar{x}_{\alpha}}-
{\partial V_{\alpha} \over \partial \bar{x}_{\alpha}} \cdot{\partial f^{(N)}
\over \partial
\bar{v}_{\alpha}} \right]=0 \eqno(53)$$
where the potential $ V_{\alpha}$ seen by particle $\alpha$ can be written as a
sum of two-body potentials, viz.
$$V_{\alpha}=\sum_{\beta \neq \alpha}\phi_{\alpha \beta} \eqno(54)$$
\noindent
If the distribution function $f^{(N)}(\bar{w}_{\alpha},t)$ is both symmetric
under exchange
of any of the particles and sufficiently convergent asymptotically for all
$\bar{w}_{\alpha}$,
it then follows upon integrating Eq. (53)
(see e.g. Binney and Tremaine 1987) that the one and two particle
distribution functions
$$f^{(1)}(\bar{w}_{1},t)=\int f^{(N)}d^6\bar{w}_2...d^6\bar{w}_N$$
$$f^{(2)}(\bar{w}_{1},\bar{w}_{2},t)=\int f^{(N)}d^6\bar{w}_3...d^6\bar{w}_N
\eqno(55)$$
are related via
$${\partial f^{(1)} \over \partial t}+
\bar{v}_{1} \cdot {\partial f^{(1)} \over \partial \bar{x}_{1}}=
(N-1)\int {\partial \phi_{12} \over \partial \bar{x}_{1}} \cdot
{\partial f^{(2)} \over \partial \bar{v}_{1}}d^6\bar{w}_2
\eqno(56)$$
\noindent
In terms of the two-particle correlation function defined by
$$g(\bar{w}_1,\bar{w}_2,t)=f^{(2)}(\bar{w}_1,\bar{w}_2,t)-
f^{(1)}(\bar{w}_1,t)f^{(1)}(\bar{w}_2,t) \eqno(57)$$
and the conventional kinetic theory phase space density $f(\bar{w}_1,t)=
Nf^{(1)}(\bar{w}_1,t)$ which is normalized according to
$$\int f(\bar{x},\bar{v},t)d^3\bar{v}=\sigma(\bar{x},t)~~~,~~~\int
\sigma(\bar{x},t)d^3\bar{x}=N \eqno(58)$$
we find that Eq. (56) then reduces (for large $N$) to
$${\partial f(\bar{x},\bar{v},t) \over \partial t}+
\bar{v} \cdot {\partial f(\bar{x},\bar{v},t) \over \partial \bar{x}}-
{\partial V(\bar{x}) \over \partial \bar{x}} \cdot {\partial
f(\bar{x},\bar{v},t)
\over \partial \bar{v}}$$
$$=N^2 {\partial \over \partial \bar{v}} \cdot
\int {\partial \phi(\bar{x},\bar{x}_{2}) \over \partial \bar{x}}
g(\bar{x},\bar{v},\bar{x}_{2},\bar{v}_{2},t)d^3\bar{x}_2d^3\bar{v}_2
\eqno(59)$$
where
$$V(\bar{x})=\int
\phi(\bar{x},\bar{x}_2)f(\bar{x}_2,\bar{v}_2,t)d^3\bar{x}_2d^3\bar{v}_2
=\int \phi(\bar{x},\bar{x}_2)\sigma(\bar{x}_2,t) d^3 \bar{x}_2\eqno(60)$$
with Eq. (59) actually being exact in the large $N$ limit.
While the above derivation is completely
standard, our point in repeating it here is to bring out the fact that there is
no need to specify
the explicit form of the potential in order to derive  Eq. (59), with Eq. (59)
thus
still being valid even in the presence of our linear potential. We thus extend
BBGKY (as opposed to the collisionful Boltzmann equation) to include
linear potentials.

In order to extract out some general information from Eq. (59) we note first
that if we set the two-body correlation  $g(\bar{w}_1,\bar{w}_2,t)$ to zero,
Eq. (59)
reduces to the Vlasov equation
$${\partial f(\bar{x},\bar{v},t) \over \partial t}+
\bar{v} \cdot {\partial f(\bar{x},\bar{v},t) \over \partial \bar{x}}-
{\partial V(\bar{x}) \over \partial \bar{x}} \cdot {\partial
f(\bar{x},\bar{v},t)
\over \partial \bar{v}}=0
\eqno(61)$$
which is often used in discussions of clusters, with $V(\bar{x})$ of Eq. (60)
then serving as the self-consistent mean gravitational field in which each
particle moves. (Equation (61) is often also called the collisionless Boltzmann
equation though in general there are in fact some differences between the
Vlasov and collisionless Boltzmann equations which we comment on briefly below,
but, independent of what name it may be given, the important point is that we
are only able to use Eq. (61) for
clusters when correlations are negligible).  However, suppose we do
not in fact drop the two-body correlation term in Eq. (59)
but instead continue to carry it. Then, since its right hand side is a total
divergence with respect to velocity, if we integrate
Eq. (59) over $d^3\bar{v}$ the correlation term will simply make no
contribution in this
integration. (We assume here
and throughout that the correlation term falls fast enough to compensate for
the growing linear
potential so that the surface terms are in fact negligible). Moreover, the
velocity derivative term
on the left hand
side of Eq. (59) would similarly integrate away in this case, and on
introducing the
one-particle distribution averages (here and throughout `$<~>$' will mean
averaged with respect to
the velocity distribution)

$$\sigma(\bar{x},t)<v_i>=\int v_i f(\bar{x},\bar{v},t)d^3\bar{v}$$
$$\sigma(\bar{x},t)<v_iv_j>=\int v_iv_j f(\bar{x},\bar{v},t)d^3\bar{v}~~~,~~~
P_{ij}=<v_iv_j>-<v_i><v_j> \eqno(62)$$
($i=1,2,3$), we therefore obtain
$${\partial \sigma(\bar{x},t) \over \partial t} +
{\partial  \over \partial \bar{x}}\cdot \left(\sigma(\bar{x},t)<\bar{v}>
\right)=0
\eqno(63)$$
which we recognize as the standard kinetic theory continuity equation.
Further, if we multiply Eq. (59) by $v_i$ first
and then integrate over all velocity, we next obtain
$${\partial \over \partial t}\left(\sigma(\bar{x},t)<v_i>\right)+
{\partial \over \partial x_j} \left(\sigma(\bar{x},t)<v_iv_j>\right)+
\sigma(\bar{x},t){\partial V(\bar{x}) \over \partial x_i}$$
$$=\sigma(\bar{x},t){\partial <v_i>  \over \partial t}  +
\sigma(\bar{x},t)<v_j>{\partial <v_i> \over \partial x_j}+
{\partial \over \partial x_j} \left(\sigma(\bar{x},t)P_{ij} \right)+
\sigma(\bar{x},t){\partial V(\bar{x}) \over \partial x_i}
$$
$$=-N^2 \int {\partial \phi(\bar{x},\bar{x}_{2}) \over \partial x_i}
g(\bar{x},\bar{v},\bar{x}_{2},\bar{v}_{2},t)
d^3\bar{v}d^3\bar{x}_2d^3\bar{v}_2   \eqno(64)$$
following an integration by parts and the use of Eq. (63). Equation (64) thus
differs from the
familiar Jeans or Euler equation by
virtue of the presence of the two-body correlation term on the right hand side.
Finally, if we
also contract with $x_i$ and then integrate over all $\bar{x}$ we obtain
$${\partial \over \partial t} \left(
\int \sigma(\bar{x},t)<\bar{x} \cdot \bar{v}> d^3 \bar{x}\right)-
\int \sigma(\bar{x},t)<v^2> d^3 \bar{x}+
\int \sigma(\bar{x},t)<\bar{x} \cdot
{\partial V(\bar{x}) \over \partial \bar{x}}>d^3 \bar{x}$$
$$=-N^2
\int \bar{x} \cdot {\partial \phi(\bar{x},\bar{x}_{2}) \over \partial \bar{x}}
g(\bar{x},\bar{v},\bar{x}_{2},\bar{v}_{2},t)
d^3\bar{x}d^3\bar{v}d^3\bar{x}_2d^3\bar{v}_2 \eqno(65)$$
which would recover the virial relation of Eq. (50) only if we were to set the
correlation term to zero.

Since the virial relation of Eq. (50) is generally regarded as being an exact
relation, it is necessary to explain why we are only able to recover it from
Eq. (65) if there are no correlations. To this end, if we return to the exact
BBGKY equation of Eq. (56) and proceed to average it just as we averaged Eq.
(59), then the continuity equation of Eq. (63) would still obtain, but
Eqs. (64) and (65) would respectively be replaced by
$${\partial \over \partial t}\left(\sigma(\bar{x},t)<v_i>\right)+
{\partial \over \partial x_j} \left(\sigma(\bar{x},t)<v_iv_j>\right)
$$
$$=-N^2 \int {\partial \phi(\bar{x},\bar{x}_{2}) \over \partial x_i}
f^{(2)}(\bar{x},\bar{v},\bar{x}_{2},\bar{v}_{2},t)
d^3\bar{v}d^3\bar{x}_2d^3\bar{v}_2   \eqno(66)$$
and
$${\partial \over \partial t} \left(
\int \sigma(\bar{x},t)<\bar{x} \cdot \bar{v}> d^3 \bar{x}\right)-
\int \sigma(\bar{x},t)<v^2> d^3 \bar{x}$$
$$=-N^2
\int \bar{x} \cdot {\partial \phi(\bar{x},\bar{x}_{2}) \over \partial \bar{x}}
f^{(2)}(\bar{x},\bar{v},\bar{x}_{2},\bar{v}_{2},t)
d^3\bar{x}d^3\bar{v}d^3\bar{x}_2d^3\bar{v}_2 \eqno(67)$$
which involve two-particle averages of the potential. Moreover, if the two-body
potential has a power law dependence $\phi_{12}=|\bar{x}_1-\bar{x}_2|^n$, we
can
then reexpress Eq. (67) as
$${\partial \over \partial t} \left(
\int \sigma(\bar{x},t)<\bar{x} \cdot \bar{v}> d^3 \bar{x}\right)-
\int \sigma(\bar{x},t)<v^2> d^3 \bar{x}$$
$$=-{nN^2 \over 2}
\int \phi(\bar{x},\bar{x}_{2})
f^{(2)}(\bar{x},\bar{v},\bar{x}_{2},\bar{v}_{2},t)
d^3\bar{x}d^3\bar{v}d^3\bar{x}_2d^3\bar{v}_2 \eqno(68)$$
a completely exact relation which should be regarded as the true virial in
which
the two-particle potential is averaged with the full two-body distribution thus
allowing for the possibility that the particles can be correlated. In fact only
if particles are uncorrelated does Eq. (68) reduce to Eq. (50) with the
correlation energy then dropping out. Usually in classical mechanics we define
the
potential energy of an $N$ particle system to be that exhibited in Eq. (50).
However, that energy is the work done in independently bringing each particle
in
from infinity one by one in the mean field provided by those that had already
been
brought in, and thus simply ignores any possible correlations, i.e. even though
the insertion of
the potential $V_{\alpha}$ of Eq. (54) into Eq. (49) generates a two-body sum,
by construction
that sum is an uncorrelated one to thus yield the one-body
Eq. (50). (In passing we note since the insertion of Eq. (54) into our
starting point of Eq. (47) does not generate any such double sum, the equation
of
motion itself is always only a one-body equation independent of whether or not
there are
any two-body correlations - indeed the correlations of the BBGKY equations are
generated
statistically starting from the correlation insensitive Eq. (47). Now
equations of motion such as Eq. (47) can also be derived starting from a
Lagrangian (or
a Hamiltonian) which involves the total potential energy written as the
uncorrelated double
sum $\Sigma \phi(\alpha, \beta)/2$ over all $\{\alpha,~\beta \neq \alpha\}$.
However that double
sum can only serve as the potential energy when there are no correlations, and
even though its
(uncorrelated) Euler-Lagrange variation does indeed lead to Eq. (47), Eq. (47)
is also valid even
in the presence of correlations when it must instead be associated with Eq.
(68), to
thereby account in general for both equilibrium and non-equilibrium situations.
In
classical kinetic theory then the equation of motion and the Liouville operator
of Eq. (53)
thus have primacy over the Lagrangian which is only readily
definable for uncorrelated (equilibrium) systems.) Thus in the presence of
correlations it is
Eq. (68) rather than Eq. (50) that should be used in general, with Eq. (50)
only emerging
at times late enough that all correlations have had time to die out, a steady
state
situation in which the system is conventionally referred to as being
virialized.
Hence we see that in the presence of correlations we must not merely not use
Eq.
(50), but we also should not expect $\ddot{I}$ to be zero. Moreover, since a
steady state solution to the correlationless Vlasov equation would only mildly
constrain the one-particle distribution function by requiring it to be a
function
only of the total single particle energy $mv^2/2+mV(\bar{x})$ (and also the
angular
momentum if the system is not isotropic), we see that the specific
dependence on energy which eventually will emerge at late times for any given
system must be fixed by the manner in which the correlations approach zero at
late times, something which can only be fixed by the entire BBGKY all order
hierarchy, with the next equation in the hierarchy for instance being
$${\partial f^{(2)} \over \partial t}+
\bar{v}_{1} \cdot {\partial f^{(2)} \over \partial \bar{x}_{1}}
+\bar{v}_{2} \cdot {\partial f^{(2)} \over \partial \bar{x}_{2}}
-{\partial \phi_{12} \over \partial \bar{x}_{1}}\cdot \left(
{\partial f^{(2)} \over \partial \bar{v}_{1}}-
{\partial f^{(2)} \over \partial \bar{v}_{2}} \right)$$
$$=(N-2) {\partial  \over \partial \bar{v}_{1}} \cdot
\int {\partial \phi_{13} \over \partial \bar{x}_{1}}f^{(3)}d^6\bar {w}_3 +
(N-2) {\partial  \over \partial \bar{v}_{2}} \cdot
\int {\partial \phi_{23} \over \partial \bar{x}_{2}}f^{(3)}d^6\bar {w}_3
\eqno(69)$$
in an essentially intractable infinite chain. Thus we see that the
two-body correlations have to play a crucial role in the evolution of the
system
which imprints itself on the ultimate late time equilibrium distribution
function (assuming that equilibrium ever occurs) in an essentially completely
unknown way
for which there is currently no clear guidance.

Before leaving the general discussion of the BBGKY hierarchy, it is of some
interest to
compare the BBGKY approach with that of the standard Boltzmann equation
approach.
In kinetic theory the great practical difficulty in using the BBGKY hierarchy
is that knowledge
of the form of the two-body distribution is needed in order to determine the
one-body distribution
$f(\bar{x}, \bar{v},t)$. As an alternative to this chain we could instead
consider the master
equation which explicitly counts how many particles enter and leave a given
region of phase space
through two-body collisions. Thus in general if particles are sitting in some
global external
potential $V_{ext}(\bar{x})$ (i.e. truly external to the system of particles of
interest) and
undergo local scattering in and out of some region of phase space $\bar{w}$
through two-body
interparticle collisions, then in general we may write (e.g. Binney and
Tremaine 1987)
$${\partial f(\bar{x}, \bar{v},t) \over \partial t}+
\bar{v} \cdot {\partial f(\bar{x},\bar{v},t) \over \partial \bar{x}}-
{\partial V_{ext}(\bar{x}) \over \partial \bar{x}} \cdot {\partial
f(\bar{x},\bar{v},t)
\over \partial \bar{v}}$$
$$=\int [\Psi(\bar{w} -\Delta \bar{w},\Delta \bar{w}) f(\bar{w} -\Delta
\bar{w})
- \Psi(\bar{w},\Delta \bar{w})f(\bar{w})]
d^6 \Delta\bar{w} \eqno(70)$$
where $\Psi(\bar{w},\Delta \bar{w})d^6 \Delta\bar{w}$ is the probability per
unit time that
particles with coordinates $\bar{w}$ will scatter into phase space volume $d^6
\Delta\bar{w}$.
As derived, Eq. (70) requires the explicit splitting of the potential of Eq.
(47) into clearly
distinguishable external and scattering pieces (which typically in standard
kinetic theory would
mean some external electromagnetic or gravitational field applied to a gas
which experiences local
molecular interactions). While exact, Eq. (70) is just as intractable as Eq.
(59) since the two-body
distribution is again involved this time through the presence of the two-body
scattering
probability $\Psi(\bar{w},\Delta \bar{w})$. In order to proceed, some
approximation needs to be made
in which the two-body distribution can be expressed in terms of $f(\bar{x},
\bar{v},t)$ in some way.
Boltzmann's own approach was to begin with the master equation, restrict to
local (molecular)
two-body scattering and proceed from the assumption of molecular chaos.
Alternative approaches which
work down from the BBGKY equations and lead to similar results in the
short-range molecular case
are described in Huang (1987) and Liboff (1990), with all of these approaches
leading to the
collisionful Boltzmann equation for the one-particle distribution function,
viz.
$${\partial f(\bar{x}, \bar{v},t) \over \partial t}+
\bar{v} \cdot {\partial f(\bar{x},\bar{v},t) \over \partial \bar{x}}-
{\partial V_{ext}(\bar{x}) \over \partial \bar{x}} \cdot {\partial
f(\bar{x},\bar{v},t)
\over \partial \bar{v}}$$
$$=\int  |\bar {v}-\bar {v}_2| \sigma (\Omega)
(f(\bar{x}, \bar{v}^{\prime},t)f(\bar{x},\bar{v}_2^{\prime},t)
-f(\bar{x}, \bar{v},t)f(\bar{x}, \bar{v}_2,t)) d^3\bar {v}_2 d\Omega\eqno(71)$$
written here quite generally for a system of particles in an external potential
$V_{ext}(\bar{x})$
undergoing local momentum conserving two-body collisions $\bar{v}+\bar{v}_2
\rightarrow
\bar{v}^{\prime}+\bar{v}_2^{\prime}$ through scattering angle $\Omega$  with
differential cross section $\sigma (\Omega)$.

For a purely self-gravitating system such as a cluster there is no explicit
external
$V_{ext}(\bar{x})$ term to begin with (since the right hand side of the master
equation would then
necessarily (by definition) have to contain the effects of all of the
gravitational scatterings
within the self-gravitating cluster). Thus the first key question to ask in
possible applications
of Eq. (71) to clusters is whether Eq. (71) (with $V_{ext}(\bar{x})=0$)
actually follows from the
BBGKY hierarchy or master equation at all. (If it were to do so, then the
collisionless Boltzmann
equation would technically then simply be $\partial f / \partial t+ \bar{v}
\cdot \partial f/
\partial \bar{x} =0$ rather than the Vlasov equation of Eq. (61).) Apart from
the issue of the a
priori validity of Eq. (71) in the self-gravitating case, it is further
generally assumed in the
literature that in the self-gravitating case the total gravitational potential
of Eq. (47) may
actually be divided up into separate local and global contributions. Even
though it is not
immediately clear how this can be done in general (and certainly not clear in
cases with long range
potentials which grow with distance), and despite the fact that it might even
involve a double
counting problem, $V_{ext}(\bar{x})$ is usually identified with the mean
gravitational field
$V(\bar{x})$ of Eq. (60) and Eq. (71) is replaced by
$${\partial f(\bar{x},\bar{v},t) \over \partial t}+
\bar{v} \cdot {\partial f(\bar{x},\bar{v},t) \over \partial \bar{x}}-
{\partial V(\bar{x}) \over \partial \bar{x}} \cdot {\partial
f(\bar{x},\bar{v},t)
\over \partial \bar{v}}$$
$$=\int  |\bar {v}-\bar {v}_2| \sigma (\Omega)
(f(\bar{x}, \bar{v}^{\prime},t)f(\bar{x},\bar{v}_2^{\prime},t)
-f(\bar{x}, \bar{v},t)f(\bar{x}, \bar{v}_2,t)) d^3\bar {v}_2 d\Omega
\eqno(72)$$
to yield an equation which then of course would yield the Vlasov equation where
we to then drop
the collision integral term. While Eq. (72) even appears plausible in that
it identifies $V_{ext}(\bar{x})$ with the mean gravitational field
$V(\bar{x})$, there would not
appear to be any explicit formal derivation of Eq. (72) for self-gravitating
systems in the
literature starting from either Eq. (59) or Eq. (70), even apparently in the
well-studied $1/r$
potential case. Thus the
second key issue for clusters then is not so much what the implications of Eq.
(72)
might be, but rather how valid it in fact is in the first place. However, in
passing, it is
interesting to note (see e.g. Huang 1987) that since elastic two-body
collisions conserve the total
momentum, the total energy, and the total number of particles, the integration
over all velocities
of the product of any of these conserved quantities with the collision integral
will automatically
give zero. The collision integral term thus makes no contribution in these
averagings, with
averaging of Eq. (72) thus immediately yielding (for any potential in fact for
which Eq. (72) is
valid) the correlationless version of Eqs. (63) - (65), i.e. the form these
equations would take if
the two-body correlation function $g(\bar{w}_1,\bar{w}_2,t)$ were set to zero
in them. Thus, once
we have the collisionful Boltzmann equation in the form of Eq. (72) at all, the
standard
gravitational Jeans and virial equations would follow, even in the presence of
the collision
integral term, without any  need to invoke the Vlasov equation. The Jeans
equation is thus in
principle valid even in the presence of the two-body collisions of Eq. (72),
with the Jeans
equation then holding even for distributions which do not obey the Vlasov
equation, so that the
Vlasov equation would then only be sufficient to yield the Jeans equation, but
not necessary.

To explore the issue of whether Eq. (72) is in fact actually valid in the first
place,
let us suppose for the sake of the argument that Eq. (72) is in fact valid for
Newtonian
clusters. For them the collision integral term turns out to be negligible on
small time
scales. Specifically, since the two-body cross section in a $1/r$ potential is
of order
$\sigma \sim m^2G^2/v^4$ while a typical velocity is of order $v^2 \sim NmG/R$
for an N particle
cluster with radius $R$, the collision integral term in Eq. (72) is of order
$f/Nt_c$ where
$t_c=R/v$ is the cluster crossing time. Thus only after $N$ crossing times will
the collision
integral compete with the left hand side of Eq. (72), with Eq. (72) thus
reducing to the Vlasov
equation of Eq. (61) at short times, a condition which only mildly constrains
the distribution
function by requiring it to be a function only of the total single particle
energy
$mv^2/2+mV(\bar{x})$ (should the distribution function be time independent that
is in this epoch,
with this assumption actually being an additional independent constraint beyond
simply assuming the
actual validity of the Vlasov equation itself). On the other hand, at very late
times, through the
very fact of rescattering, Eq. (72) would lead us to an actual specification of
the functional
dependence of $f(\bar{x},\bar{v},t)$ on the energy, i.e. the Maxwell-Boltzmann
distribution
$f \sim exp(-(mv^2/2+mV(\bar{x}))/kT)$ which is an exact solution to the
collisionful Eq. (72) in
which the collision integral term vanishes non-trivially through the vanishing
of
$f(\bar{v}^{\prime})f(\bar{v}_2^{\prime}) - f(\bar{v})f(\bar{v}_2)$. Thus,
unlike the collisionful
Boltzmann equation, a  collisionless Boltzmann equation simply does not contain
enough information
to fix the distribution function completely. Hence if the collision integral is
indeed numerically
small in some kinematic regime, the distribution function which eventually
results in that regime
must be fixed by something else, with the only apparent other candidate being
the multi-body
correlations of the entire all order BBGKY hierarchy, in which case Eq. (72)
could not have been
valid at that time in the first place. Or stated differently, if the Vlasov
equation is valid at
early rather than late times, then nothing is available to fix the form of the
distribution
function at that earlier time. Thus the validity of the collisionful Eq. (72)
for Newtonian
clusters would require the
following somewhat peculiar time development profile. First, at some very early
time even prior to
the onset of Eq. (72) the two-body correlation term in Eq. (59) would have to
be important. Then it
would have to gradually dampen to zero as it forces the system into some
particular solution to the
Vlasov equation of Eq. (61) with some particular dependence on the energy then
being determined
(if $\partial f(\bar{x},\bar{v},t)/ \partial t$ is in fact zero in this
regime). Then at later times
still the collision integral term effects would have to become important (even
though the two-body
correlation $g(\bar{w}_1,\bar{w}_2,t)$ would still be required to be negligible
because of the
continuing imposition of molecular chaos) and move the system away from being
in a solution to the
Vlasov equation, and then finally at the very latest times the system would
then have to thermalize
into Maxwell-Boltzmann. There might thus appear to be some difficulties for
clusters if the
collisionful Boltzmann equation is ever valid in the self-gravitating case,
with the smallness of
the right hand side of Eq. (72) apparently turning out to be something of a
minus rather than a
plus for Newtonian clusters. On the other hand, there would not appear to be
any formal difficulty
in either the Newtonian or linear potential case in having the cluster evolve
directly via BBGKY
into a late time Vlasov equation through the late time vanishing of the
correlation functions in a
time development which imprints itself on the resulting distribution function
while never passing
through any Boltzmann equation regime at all. Thus we shall restrict our
discussion of clusters to
the use of the BBGKY hierarchy without regard to the Boltzmann equation at all.
(Of course, in
practice if one only uses the Vlasov equation of Eq. (61), it does not
particularly matter whether
it came from BBGKY or from the collisionful Boltzmann equation anyway. It would
only matter if one
wants to follow the approach to equilibrium of a not yet fully virialized
system). While there may
not be any formal difficulty in using our preferred choice of the BBGKY
hierarchy, there is of
course a great practical difficulty though, since without knowledge of the time
development of the
correlation functions, it is not be possible to ascertain into which particular
distribution
function the system eventually does develop in any such late time Vlasov
equation regime. Because
of this, we shall below only seek implications of the BBGKY approach which
require no knowledge of
the explicit form of the one-particle distribution function (a procedure which
incidentally releases
us from needing to invoke any of the popular model distribution functions often
considered in
cluster studies, models for which the literature seems not to present any a
priori justification).

While a first principles evaluation of the validity of the virial must await a
determination of the
two-body correlation function, it is possible to establish some
phenomenological expectations using
observed cluster data. For the most studied cluster, the Coma cluster, the
relevant data may be
found in Kent and Gunn (1982), The and White (1986), and White et al (1993). At
the distance of Coma
an arc minute is $20/h$ kpc (for a Hubble parameter $H_0 =100h$ km/sec/Mpc), so
that the standard
Abell radius is $75^{\prime}$. The surface brightness may be approximately
fitted (see below) by a
modified Hubble profile ($\sim 1/(R^2+R_0^2)$) with a core radius
$R_0=9.23^{\prime}$, and the
observed cluster data go out about to $3^{\circ}\simeq 20R_0$ or so. (It is not
immediately clear
where the cluster actually ends, a point we examine below.) White et al (1993)
quote a
total blue surface luminosity within the Abell radius of $L_B=1.95 \times
10^{12}/h^2~L_{B\odot}$,
and a mean projected line of sight velocity of 970 km/sec for a convenient
magnitude limited cut on
the data which restricts to $R\leq 120^{\prime}$ (the revised binning of The
and White to the
original velocity data of Kent and Gunn is reproduced here as Fig. (3)). For
such a mean velocity,
the time required to cross the associated $240^{\prime}$ diameter is $1.5
\times 10^{17}/h$ sec,
which is of order $1/2H_0$, i.e. of order half a Hubble time, and thus we
should not expect the
entire cluster to have yet had time to virialize. Hence, for the purposes of
this study, we shall
simply assume that only the inner region cluster core  - a region which still
turns out to contain a
sizable fraction of the entire cluster - has so far virialized. (While we can
readily assert that we
would not expect virialization after only one crossing time, we note in passing
that having a time
long enough for quite a few  crossing times is not in and of itself sufficient
to ensure that we
then would have virialization, since the relevant time scale is not the
crossing time but the BBGKY
correlation  relaxation time, a time which is currently unknown; and in general
it would appear to
require a somewhat subjective judgment to say exactly just how big a fraction
of the entire cluster
has yet had time to virialize.) Further, the very fact that the cluster surface
brightness does fall
so slowly - unlike the rapid, exponential drop within the much smaller
individual spiral galaxies -
could also be an indicator that the entire cluster has not in fact yet had
enough time to have
virialized completely by compactifying itself into a relatively small volume
with a much more
steeply declining surface brightness (or to have yet succeeded in decoupling
itself from the
background provided by the rest of the Universe either for that matter).
Moreover, while the
potentials in the conformal theory grow with distance, tidal forces will still
fall (linearly) with
distance and be proportional to the ratio of the size of orbit being perturbed
to the distance
between the two systems of interest. Thus while the rotation curves of
individual spiral galaxies
are essentially unaffected by the presence of any nearby galaxy, clusters in
any nearby supercluster
would, given their huge amount of matter, be expected to have some influence on
the larger orbits of
a given cluster of interest while only marginally affecting the smaller orbits,
so again
virialization is more reasonable for the core than for the periphery.

Before we proceed to get an actual numerical estimate for the virial velocity
in the conformal
gravity theory, we note first that since, as can be seen from Fig. (3), the
projected line of sight
velocity falls with distance, this might immediately appear to exclude any
theory in which the
potential grows rather than falls with distance. However, it turns out that the
projected velocity
can still actually fall in our theory simply because of the difference between
the full
three-dimensional and the projected two-dimensional averaging procedures. For a
spherically
symmetric system the projected line of sight distribution average
$<\sigma_p^2(R)>$, where $R$ is the impact parameter, is related to the
previously introduced
three-dimensional radial and tangential velocity distribution averages
according to
$$I(R)<\sigma_p^2(R)>=2\int_R^{\infty} {dr \sigma(r) \over r(r^2-R^2)^{1/2}}
\left(
<v_r^2>(r^2-R^2)+ <v_{\theta}^2>R^2 \right)
\eqno(73)$$
\noindent
(The surface mass density $I(R)$ is related to the volume mass density
$\sigma(r)$ via
Eq. (26).) For a system which has already virialized, we may drop both the
correlation term
and the explicit average velocity time derivative term from Eq. (64), with a
spherically symmetric
steady state Jeans equation then yielding
$${d \over dr}(\sigma(r)<v_r^2>)+{2 \sigma(r) \over r}(<v_r^2>-<v_{\theta}^2>)
=-\sigma(r)V^{\prime}(r)
\eqno(74)$$
\noindent
While it is not possible to solve Eqs. (73) and (74) in a closed form without
further input, it
is possible to extract out some general features from a study of some simple
cases. If for instance
we assume that the system is isotropic (so that $<v_r^2>=<v_{\theta}^2>$), Eq.
(73) then yields
a closed form for the dependence of $<\sigma_p^2(R)>$ on the potential, viz.
$$I(R)<\sigma_p^2(R)>=2\int_R^{\infty}dr \sigma(r) (r^2-R^2)^{1/2}V^{\prime}(r)
\eqno(75)$$
\noindent
Similarly, if we assume that the system is purely circular ($<v_r^2>=0$), we
obtain
$$I(R)<\sigma_p^2(R)>=R^2\int_R^{\infty}dr {\sigma(r) V^{\prime}(r) \over
(r^2-R^2)^{1/2}}
\eqno(76)$$
while a purely radial system ($<v_{\theta}^2>=0$) yields
$$I(R)<\sigma_p^2(R)>={1 \over R}\int_R^{\infty}dr \sigma(r) V^{\prime}(r)
\left( r^2 arcsin \left( (1-{R^2 \over r^2})^{1/2} \right) -R(r^2-R^2)^{1/2}
\right)
\eqno(77)$$
\noindent
Thus the radial case will tend to emphasize small $R$, the circular case large
$R$,
and the isotropic case intermediate $R$.

To see how this works out in practice we can evaluate Eqs. (75) - (77) for the
illustrative case
of a cluster with a modified Hubble surface profile.  Since such a slowly
falling surface density
would yield an infinite number of particles if integrated to infinity, we must
cut off the density
at some maximum radius $R_M$. However, the very act of cutting off the surface
brightness has some
explicit consequences for the Abel transform of Eq. (26). If, for instance, we
simply give the
surface brightness the sharp cut-off $I(R) \theta (R-R_M)$, its insertion into
the Abel transform
would then induce an unphysical singular term $-I(R_M)/\pi (R_M^2-r^2)^{1/2}$
in the volume
density coming from the delta function derivative of the theta function. (This
singularity is needed
to recover $I(R)$ when integrating back over the volume density since the range
$R$ to $R_M$
for the volume density integration shrinks to zero as $R$ approaches $R_M$.)
However,
there is no such problem in giving the volume density a sharp cut-off instead,
since the generic
matter volume density
$$\sigma(r)={\sigma_0  \over (r^2+R_0^2)^{3/2}} \theta (R_M-r)
\eqno (78)$$
\noindent
then yields
$$I(R)={2\sigma_0 (R_M^2-R^2)^{1/2} \over (R^2+R_0^2)(R_M^2+R_0^2)^{1/2}}
 \theta (R_M-R)
\eqno (79)$$
as the generic matter surface density, a surface density which now nicely
vanishes smoothly rather
than sharply at the cut-off, and consequently in our explicit numerical fitting
to the Coma cluster
to be presented below we shall actually use Eq. (79) as a model for the surface
brightness rather
just the plain $2\sigma_0/(R^2+R_0^2)$. Given the generic Eqs. (78) and  (79)
it is now
straightforward to evaluate the various line of sight velocities for both the
Newtonian
and the linear potential cases via the direct use of Eq. (25), and we present
the resulting velocity
curves (calculated with $R_M=20R_0$ for explicitness) as Figs. (4) and (5)
respectively.
As can be seen, in the Newtonian case all the three discussed possibilities
have their maxima at small radii (the purely circular case is actually
asymptotically
flat in projection), while the linear potential case
shows a radically different behavior. Specifically, the pure radial case peaks
at very small $R$,
the pure isotropic case at $R_M/2$, and the pure circular case at $R_M$. Thus,
depending
on the radial to tangential velocity ratio, in fully virialized clusters it is
quite possible for
the projected line of sight velocity to fall at large distances even in a
theory with rising
potentials. (Essentially at large impact parameters the amount of cluster
material along a line of
sight goes to zero faster than the rate at which the potential grows).
We thus identify a somewhat unusual projection effect, and see that in general
the
curves of Figs. (4) and (5) could eventually turn out to be very useful in
discriminating
between Newtonian and linear potentials in systems which are in fact fully
virialized.

Turning now to clusters which have not yet had time to virialize completely, we
first evaluate
the fractional amount of matter contained in the core region. For the densities
of Eqs. (78)
and (79) we can readily evaluate the fractional amount of matter $4\pi \int
dr^{\prime}r^{\prime 2}
\sigma (r^{\prime})/N$ within a given volume of radius $r$ and the fractional
amount of matter
$2 \pi \int dR^{\prime}R^{\prime}I(R^{\prime})/N$ within a given projected
surface of impact
parameter $R$. As we can see from the curves of Fig. (6) (which are calculated
with $R_M=20R_0$ for
explicitness), one quarter of the matter by volume is
contained within $r<2.5R_0$ and one half within $r<5R_0$, while one half of the
matter by surface is
contained within $R<4R_0$. The central region of the cluster thus contains a
sizable portion of
the entire amount of matter in the cluster, so that a virialization of only the
inner region of the
cluster is not insignificant. From the data we have only the projected
two-dimensional line of
sight velocity $<\sigma^2_p(R)>$ which is inconveniently related in Eq. (73) to
an integration of
the three-dimensional radial and tangential velocities (the ones we actually
calculate by the
virial) over both the inner and outer regions of the cluster. However, an
integration of Eq. (73)
itself over a sphere of radius $r_{m}$ then crucially projects out the
undesired
unvirialized $r > r_m$ region from the $R \leq r \leq R_M$ integration range
involved in Eq. (73)
to yield
$$2 \pi \int_{0}^{r_{m}} dR RI(R)<\sigma _p^2(R)>$$
$$=2\pi \int_{0}^{r_{m}} dr r^2 \int_{0}^{\pi} d \theta sin\theta
\sigma(r) \left(
<v_r^2>cos^2\theta + <v_{\theta}^2>sin^2 \theta \right)$$
$$={4\pi \over 3}  \int_{0}^{r_m}drr^2\sigma(r) \left(<v_r^2>+2<v_{\theta}^2>
\right)
\eqno(80)$$
a relation which is actually completely general and which involves no
assumptions regarding the
relative strengths of the mean square radial and tangential velocities. From
Eq. (80) we see that
in general the averaged squared
line of sight velocity is thus one third of the averaged squared three
dimensional velocity. To
relate the right hand side of Eq. (80) to the potential we appeal to Eq. (64)
and note that if the
two-body correlation term vanishes for $r<r_{m}$ (and if $\partial
<v_i>/\partial t =0$ of course),
then the steady state Jeans equation of Eq. (74) will hold for $r<r_{m}$.
Multiplying Eq. (74) by
$r^3$ and then integrating enables to express the spatial average
$(\sigma^2_p(r_{m}))_{av}$ for
the region $r<r_{m}$ purely in terms of virialized region quantities alone
according to our
final, key relation
$$2 \pi(\sigma_p^2(r_m))_{av} \int_{0}^{r_m}dR RI(R)=
{4\pi \over 3} \int_0^{r_m}drr^2\sigma(r)rV^{\prime}(r)
\eqno(81)$$
\noindent
(The integration of $r^3$ times Eq. (74) involves surface terms at $r=0$ and
$r=r_{m}$. The
one at zero vanishes kinematically (since $<v^2_r>$ vanishes no faster than
$1/r^2$ according to
Eq. (74)), while the other one leads to a dependence on $<v^2_r(r_{m})>$. Our
assumption
that the $r<r_{m}$ region is virialized requires the vanishing of this term in
precisely the
manner which led us to the partial virial of Eq. (52) which we discussed
above.). For our
illustrative volume density of Eq. (78) (with typical cut-off of $R_M=20R_0$
for the matter
distribution) the local virials associated with Eq. (81) are readily
evaluated, with the partial virial spatially averaged root mean square
projected velocity
$\sigma_p(r_m)=((\sigma_p^2(r_m))_{av})^{1/2}$ being plotted as a function of
$r_{m}$ in Fig. (7)
for both the Newtonian and linear cases. In Fig. (7)
the Newtonian virial is normalized to $(N\beta_{gal} c^2/R_0)^{1/2}$ while the
linear
virial is normalized to $(N\gamma_{gal} c^2R_0)^{1/2}$ where $\beta_{gal}$ and
$\gamma_{gal}$ are
individual galactic potential parameters and $N$ is the total number of
galaxies contained in the
entire cluster - and not just the number $N(r_m)$ contained in $r\leq r_m$.
Since the Newtonian
potential contribution of Eq. (25) is only sensitive to the matter interior to
$r_m$, the Newtonian
local virial very quickly levels off, but since the linear potential also feels
the matter exterior
to $r_m$, its associated local virial velocity continues to rise all the way to
$r=R_M$. Thus the
linear potential case is far more sensitive to how much of the cluster is
virialized
than the Newtonian one. (Thus in passing we note that since the Newtonian
virial is so insensitive
to how much of the cluster has in fact virialized, an application of Eq. (81)
to the core
will give predictions which are extremely close to those made under the
assumption that the entire
cluster is virialized, with the use of Eq. (81) in the standard Newtonian
theory thus essentially
being immune to the issue of how much of the cluster has in fact virialized.)

In order to apply the virial of Eq. (81) to the Coma cluster, we need first to
study the
implications of using the cut-off $I(R)$ of Eq. (79) as a model for the surface
brightness of Coma.
Rather than fit this $I(R)$ to the surface brightness, we instead opted to fit
$RI(R)$ (this being
the quantity which actually appears in the virial equations) to $R$ times the
surface brightness so
as to ensure the correct overall normalization. (The core virial velocities we
obtain below turn out
to be insensitive to this prescription). We can then fit the Coma surface
brightness data with
$R_0=9.23^{\prime}$, $R_M=20R_0=185^{\prime}$, and normalization $\sigma _0 /
R_0^3=0.016$ galaxies per cubic arc
min. Giving each galaxy an average blue luminosity of
$5.99\times10^{9}/h^2~L_{B\odot}$, then yields
the requisite total $1.95 \times 10^{12}/h^2 ~L_{B\odot}$ surface blue
luminosity within the Abell
radius, to thus fully specify $I(R)$. Using as typical the mass to light ratio
$M/L_B=5.6hM_{\odot}/L_{B\odot}$ which we obtained for the galaxy NGC 3198 (we
adjust here for the
fact that the fits of Fig. (1) were based on data which were obtained using an
adopted value of
$h=0.75$ for each of the three regular spirals in our sample) enables us to
determine the mass
volume density associated with $\sigma (r)$. It is very convenient to express
this mass density in
units of the standard critical density $\rho_c=3H_0^2/8\pi G$, and we find that
$\sigma (0^{\prime})=241.5\rho_c$, $\sigma (56.8^{\prime})=\rho_c$,
$\sigma (120^{\prime})=0.11\rho_c$, and $\sigma (185^{\prime})=0.03\rho_c$. The
cluster is thus
apparently merging with the general cosmological background at no more than
$185^{\prime}$ or so,
and would be restricted to the first $57^{\prime}$ (a region which contains
57\% of the total matter
in the cluster by volume according to Fig. (6)) if the density of the Universe
is critical. Thus in
a low density Universe we would put the edge of the cluster at $185^{\prime}$,
while in a high
density one we would only consider the potentials of the first $57^{\prime}$ of
data as contributing
to the velocity dispersion, with the next $128^{\prime}$ of data then only
contributing along with
the rest of the galaxies in the Universe to the general Hubble flow. (Noting
that the conventional
estimation of the cosmological ratio $\rho/\rho_c$ is made in comoving
coordinates while our
analysis here involves the same ratio in static coordinates, our determination
of where the static
cluster actually merges with the comoving background is thus perforce only a
rough estimate.)
Since the actual density of the Universe represents one of the key unknown
issues in cosmology, we
shall calculate core virial velocities for both the high and low density
Universe cases, and
actually find below that the values that we then obtain turn out to be
insensitive to where the
cluster ends. (In a recent paper (Mannheim 1992) it was shown that the
relativistic cosmology
associated with conformal gravity possesses no flatness problem. Unlike the
standard Einstein theory
the conformal theory thus needs no inflationary era, and its cosmological
matter density is not
required to obey $\rho=\rho_c$. Given the fact that conformal gravity also
appears to be able to
eliminate the need for galactic scale dark matter, it can thus naturally
accommodate a $\rho<\rho_c$
Universe. Nonetheless, for phenomenological completeness we shall study the
conformal theory virial
velocity predictions for both high and low density Universes, with the core
velocities turning out
to be insensitive to this whole issue anyway.)

We proceed now to an actual evaluation of the virial velocities in our model.
Our above discussion
of the mass density of Coma fixes the overall normalization of the Newtonian
potential contribution
to the virial, while taking as typical the NGC 3198 gamma to light ratio of
$9.2 \times 10^{-40}h^3 /cm/L_{B\odot}$ obtained earlier then enables us to fix
the overall
normalization of the linear potential contribution as well. Thus for a Coma
cluster composed solely
of luminous matter alone, the overall normalizations $(N\gamma_{gal}
c^2R_0)^{1/2}$ and
$(N\beta_{gal} c^2/R_0)^{1/2}$ needed for Fig. (7) take respective values of
10960 km/sec and 576
km/sec for a cluster cut off at $R_M=20R_0$ ($N=425$ galaxies).
{}From Fig. (7) we thus see that in the absence of any dark matter the luminous
Newtonian
contribution to the virial is negligibly small, while, on the other hand, the
linear contribution
associated with the luminous matter is substantial. Specifically, if the entire
$R_M=20R_0$ cluster
is virialized Eq. (81) yields a virial velocity $\sigma_p(20R_0)=10178$ km/sec,
while also yielding
partial virial velocities $\sigma_p(R_0)=1089$ km/sec, $\sigma_p(1.5R_0)=1678$
km/sec,
$\sigma_p(2R_0)=2195$ km/sec, and $\sigma_p(6.15R_0)=5018$ km/sec in various
inner regions.
Similarly, if we cut off the cluster at $56.8^{\prime}=6.15R_0$ (to yield
$N=242$ galaxies,
$(N\gamma_{gal} c^2R_0)^{1/2}=8261$ km/sec, $(N\beta_{gal} c^2/R_0)^{1/2}=435$
km/sec)
we obtain the partial virial
velocities $\sigma_p(R_0)=1028$ km/sec, $\sigma_p(1.5R_0)=1583$ km/sec,
$\sigma_p(2R_0)=2070$ km/sec,
and $\sigma_p(6.15R_0)=4885$ km/sec. The core region velocities are thus
essentially insensitive to
whether we use a high or low density Universe cut-off. (This may be understood
directly from the
potential of Eq. (25), since while that potential is sensitive to points
exterior to the point of
observation, their contribution is proportional to $r^2$ which is small in the
inner core region,
to thus prevent the region outside of the core from making any substantial
contribution to
core region virial velocities). From the data points of Fig. (3) we find that
the numerical average of the first four bins of data ($R\leq 1.3R_0$) is
$1200\pm 195$ km/sec, while
that of the first five bins ($R\leq 1.7R_0$) is $1185\pm 195$ km/sec. Before we
assess the
significance of these numbers, it is important to note that once less than the
entire spherical
cluster is virialized, then any given line of sight through the sphere, even
those at small impact
parameter $R$, will pass through both virialized and non-virialized regions
(since the integral in
Eq. (73) is from $R$ all the way to the cluster cut-off $R_M$, and not merely
to the
virialization cut-off $r_{m}$), so that the detected projected velocity at that
$R$ will include
some non-virialized contributions as well. For instance, if $r \leq 2.5R_0$ is
virialized, then out
of a $20R_0$ cluster the percentage of line of sight material which involves
unvirialized radii
$r>2.5R_0$ is $25\%$ at $R=1.5R_0$, $44\%$ at $R=2R_0$, and of course $100\%$
at $R=2.5R_0$. Thus
the very use of Fig. (3) to estimate a magnitude for a virialized
$\sigma_p(r_m)$ becomes suspect
once the cluster is less than fully virialized. (In passing we note that with a
value of
$(N\gamma_{gal} c^2R_0)^{1/2}=$ 10960 km/sec, the projected curves of Fig. (5)
would far overshoot
the data for a fully virialized Coma cluster. However, even for just an inner
region virialization,
we still
could not try to fit these curves for small $R$ since even small $R$ requires a
knowledge of the
large $r$ behavior (up to $R_M$) of the cluster in the integrals of Eqs.
(75-77)). Fortunately,
however, Eq. (81) only involves integrating up to $r_{m}$, and since it
requires no knowledge of
the distribution function or of the radial to tangential velocity admixture
either, it would appear
to be far the most reliable quantity to study, especially in only partially
virialized systems.
Conformal gravity would thus appear to have no difficulty accommodating a
virialized inner cluster
region of the order of $r_{m}\sim 1.5R_0$ without needing to invoke dark
matter, and given the just
noted limitation on the use of the data of Fig. (3) in partially virialized
systems, the theory
could possibly even accommodate up to $r_{m}\sim 2.5R_0$, a region which
contains close to one
quarter by volume of all of the matter in the entire $185^{\prime}$ of the
cluster. Moreover, given
the relevant time scales which were discussed above, it would even appear to be
quite reasonable to
expect inner region virialization up to one or two scale lengths or so. While
we would certainly
not expect any larger a portion of the cluster
to have yet virialized, a first principles determination of the two-body
correlation function and of
its potential impact on Eqs. (74), (75-77), and (81) could nonetheless prove to
be very instructive,
and might possibly even turn out to be definitive for the theory. (It is also
possible to test the
conformal theory in a way which is actually insensitive to how big a fraction
of the cluster has
in fact virialized, viz. cluster gravitational lensing which responds to all
the matter in the
cluster virialized or not; thus a yet to be made study of the conformal theory
predictions for
lensing should eventually provide an independent and definitive way of testing
the theory on whole
cluster scales.) Other than this issue though, it would appear that, in the
first instance at least,
the conformal gravity theory is indeed capable of meeting the demands of
cluster virial velocity
data, with the linear potential theory thus readily being extendable from
galactic scales up
to the much larger ones associated with clusters of galaxies without
encountering any major
difficulty.

\medskip
\noindent
{\bf (5) Implications of the Microlensing Observations for Gravitational
Theory}
\medskip
With the advent of the microlensing observations of the OGLE (Udalski et al
1993, 1994), MACHO
(Alcock et al 1993) and EROS (Aubourg et al 1993) collaborations it became
possible to explore
not only whether the presumed dark matter spherical halo actually exists, but
also to address the
critical issue we raised in Sec. (3) regarding what the actual magnitude of the
mass to light ratio
of a visible disk might be. Neither microlensing off the LMC nor the Milky Way
optical searches of
the recently refurbished Hubble Space Telescope (Bahcall et al 1994, Paresce,
De Marchi and
Romaniello 1995) are so far
finding the copious amounts of conventional astrophysical dark or faint matter
that had been widely
anticipated to reside in the halo prior to these observations, while, to the
complete contrary,
microlensing off the bulge of the Galaxy is finding an unexpectedly large
number of such sources in
the plane of the Galaxy. The extreme (but not yet unequivocal) interpretation
of these data is that
there is little or no baryonic halo at all and that the inner region optical
disk is maximal with
the largest possible $M/L$ ratio, i.e. that it is precisely of noneother than
the very structure
required in Sec. (3) of the conformal gravity theory fits. (Given the fact that
the data do also
permit of some form of halo, albeit at a lower density than that favored by the
dark matter models,
we note in passing that our current lack of knowledge as to the explicit
parameters of any such halo
leaves us momentarily in the unsatisfactory position of not being able to do
precision fitting to
galactic rotation curves, in any theory of rotation curves in fact.)

Apart from not actually finding much if any of a spherical halo at all, the
general systematics
of what has in fact been found in the plane of the Galaxy now creates several
quite severe
new challenges for the standard theory. First, the very presence of all these
microlensing
sources in the plane of the Galaxy makes a Newtonian disk even less stable than
before, thus
requiring even more halo matter again to stabilize the Galaxy just at a time
when no such
halo dark matter is being found. Second, if there is still to be a halo, then
it must now be
predominantly non-baryonic and that (unlike the situation in typical dark
matter fits to dwarf
galaxies where the dark matter halo usually does contribute in the inner region
- see e.g.
Begeman, Broeils and Sanders 1991) the halo must now make no contribution
in the inner region since maximal disk luminous matter already exhausts the
velocity
there. Thus any non-baryonic dark halo must be clever enough to keep out of the
optical disk
region even as it stabilizes it, and also to normalize itself each and every
time to the luminosity
in that selfsame optical disk so as to still yield
Tully-Fisher, something which would not seem to be immediately apparent
for putative weakly interacting non-baryonic wimps.
Third, given a potentially maximal Milky Way disk, it now becomes very hard to
understand why the $M/L$ ratios of the luminous disks in dwarf galaxies should
be
as low as they have in fact been found to be in dark matter fits
(reported values are at least an order of magnitude lower than those associated
with regular spirals)
especially since the stellar populations of dwarfs are
not that different from those of regular spirals. Or stated differently, the
dwarfs now appear to have a problem not only of too little disk luminous mass
(in the outer region) but also one of
too much disk luminous mass in the inner region (which is then conveniently
finessed
by arbitrarily cutting down the associated disk mass to light ratios in the
fits, i.e. by effectively treating the matter in the disk as though it also had
a repulsive
gravitational component), with the luminous dwarf disks simply producing too
much gravity for the
dark matter fits to handle. Now it is worth noting that the dwarf galaxy
fits actually appear to fall into two categories. There are some for which a
maximal disk gives acceptable fitting (with a less than maximal one and a
consequently bigger inner region halo then just giving better fitting), and
there are some
for which maximal disks fail completely in both the shape and the normalization
of the inner region rotation curves. Thus something has to give somewhere, and
it
would therefore appear to be worthwhile to again measure the surface
brightnesses of dwarf galaxies, optimally in many filters, to
see if some optical components have been missed or if perhaps some scale length
values might change, so that the inner regions of dwarf galaxies might then in
fact be fitted with typical maximal spiral disk $M/L$ ratios after all, so that
they then
would in fact be compatible with the disk microlensing data.

As regards the conformal linear potential theory, we already noted that it
appears capable of
reproducing all the desirable aspects of galactic dark matter without needing
the dark matter
itself, and now we see that it also appears to have survived the microlensing
observations
unscathed. Thus it must indeed be regarded as viable. Given the success (so
far) of the linear
potential theory in fitting rotation curve and cluster data
without needing to invoke dark matter, it would thus appear to us that at the
present time one
cannot categorically assert that the sole gravitational potential on all
distance scales is the
Newtonian one; and that, in the linear potential, the standard $1/r$ potential
would not only
appear to have a companion but to have one which would even dominate over it
asymptotically.
Indeed, the very need for dark matter in the standard theory may simply be due
to trying to apply
just the straightforward Newtonian potential in a domain for which there is no
prior (or even
current for that matter) justification. Even though the observational
confirmation on terrestrial
to solar system distance scales of both the Newton theory and its general
relativistic Einstein
corrections technically only establishes the validity of the Newton-Einstein
theory on those scales,
nonetheless, for most workers in the field, it seems to have established the
standard theory on all
other distance scales too; despite the fact that many other theories could
potentially have the
same leading perturbative structure on a given distance scale and yet differ
radically elsewhere.
Since we have shown that the conformal theory also appears to be able to meet
the constraints of
data, one has to conclude that at the present time the Newton-Einstein theory
is only sufficient to
describe data, but not yet necessary. Indeed, it is the very absence of some
principle which would
single out the Einstein theory from amongst all other possible covariant
theories which one could
in principle at least consider which prevents the Einstein theory from yet
being a necessary
theory of gravity. In fact, in a sense, it is the absence of some underlying
principle
which would ensure its uniqueness that is the major theoretical problem for the
Einstein theory,
rather than its phenomenological inability to fit data without invoking dark
matter; with this very
lack itself actually opening the door to other contenders (Mannheim 1994).

To conclude this paper we would like to state that since the great appeal of
Einstein gravity
is in its elegance and beauty, using an approach as ad hoc and contrived as
dark matter for it
almost defeats the whole purpose, and would even appear to be at odds with
Einstein's own view of
the way nature works. Indeed, Einstein always referred to the Einstein
Equations as being a bridge
between the beautiful geometry of the Einstein tensor and the ugliness of the
energy-momentum tensor.
The dark matter idea only serves to make the energy-momentum tensor even more
ugly, and could even
be construed as a reinvention of the aether. The great
aesthetic appeal of the conformal theory is that it adds beauty to both sides
of the gravitational
equations of motion by both retaining covariance and by endowing both the sides
of the bridge with
the additional, highly restrictive, symmetry of conformal invariance; and, as
we have seen, such a
theory may even be able to eliminate the need for dark matter altogether.

The author would like to thank D. Kazanas for stimulating discussions. The
helpful comments of
N. Brandt and G. W. Fernando are also gratefully acknowledged. This work has
been supported in part
by the Department of Energy under grant No. DE-FG02-92ER40716.00.
%\vfill\eject
\medskip
\noindent
{\bf References}
\medskip
\noindent Alcock, C., Akerlof, C. W., Allsman, R. A., Axelrod, T. S., Bennett,
D. P.,Chan, S.,
Cook, K. H., Freeman, K. C., Griest, K., Marshall, S. L., Park, H.-S.,
Perlmutter, S., Peterson,
B. A., Pratt, M. R., Quinn, P. J., Rodgers, A. W., Stubbs, C. W., and
Sutherland, W. 1993,
Nature, 365, 621.
\smallskip
\noindent Aubourg, E., Bareyre, P., Brehin, S., Gros, M., Lachieze-Rey, M.,
Laurent, B.,
Lesquoy, E., Magneville, C., Milsztajn, A., Moscosco, L., Queinnec, F., Rich,
J., Spiro, M.,
Vigroux, L., Zylberajch, S., Ansari, R., Cavalier, F., Moniez, M., Beaulieu,
J.-P., Ferlet, R.,
Grison, Ph., Vidal-Madjar, A., Guibert, J., Moreau, O., Tajahmady, F., Maurice,
E., Prevot, L.,
and Gry, C. 1993, Nature, 365, 623.
\smallskip
\noindent Bahcall, J. N., Flynn, C., Gould, A., and Kirhakos, S. 1994, ApJ,
435, L51.
\smallskip
\noindent Barnaby, D., and Thronson, H. A. 1992, AJ, 103, 41.
\smallskip
\noindent Begeman, K. G. 1987, Ph.D. Thesis, Gronigen University.
\smallskip
\noindent Begeman, K. G. 1989, A\&A, 223, 47.
\smallskip
\noindent Begeman, K. G., Broeils, A. H., and Sanders, R. H. 1991,
MNRAS, 249, 523.
\smallskip
\noindent Bekenstein, J. D. 1987, The missing light puzzle: a hint about
gravitation? in Second
Canadian Conference on General Relativity and Relativistic Astrophysics, edited
by A. Coley,
C. Dyer, and T. Tupper (World Scientific Press, Singapore).
\smallskip
\noindent Binney, J., and Tremaine, S. 1987, Galactic Dynamics, Princeton
University
Press, Princeton, N. J.
\smallskip
\noindent Bosma, A. 1978, Ph.D. Thesis, Gronigen University.
\smallskip
\noindent Bosma, A. 1981, AJ, 86, 1791.
\smallskip
\noindent Carignan, C., and Freeman, K. C. 1988, ApJ, 332, L33.
\smallskip
\noindent Carignan, C., and Beaulieu, S. 1989, ApJ, 347, 760.
\smallskip
\noindent Casertano, S. 1983, MNRAS, 203, 735.
\smallskip
\noindent Casertano, S., and van Gorkom, J. H. 1991, AJ, 101, 1231.
\smallskip
\noindent Christodoulou, D. M. 1991, ApJ, 372, 471.
\smallskip
\noindent Eddington, A. S. 1922, The Mathematical Theory of Relativity,
(8th Ed. 1960; Cambridge; Cambridge University Press).
\smallskip
\noindent Freeman, K. C. 1970, ApJ, 160, 811.
\smallskip
\noindent Huang, K. 1987, Statistical Mechanics, 2nd Ed., J. Wiley, New York,
N.Y.
\smallskip
\noindent Kalnajs, A. J. 1983, in Internal Kinematics and Dynamics of Disk
Galaxies,
IAU Symposium No. 100, ed. E. Athanassoula (Reidel, Dordrecht), p. 87.
\smallskip
\noindent Kazanas, D. 1991, Astrophysical aspects of Weyl gravity, in
Nonlinear Problems in Relativity and Cosmology, Proceedings of the Sixth
Florida Workshop on Nonlinear Astronomy, University of Florida, October
1990. Edited by J. R. Buchler, S. L. Detweiler, and J. R. Ipser,
Annals of the New York Academy of Sciences, Vol. 631, 212.
\smallskip
\noindent Kazanas, D., and Mannheim, P. D. 1991a, ApJS, 76, 431.
\smallskip
\noindent Kazanas, D., and Mannheim, P. D. 1991b, Dark matter or new physics?,
in
Proceedings of the ``After the First Three Minutes" Workshop, University
of Maryland, October 1990. A. I. P. Conference Proceedings No. 222, edited by
S. S. Holt,
C. L. Bennett, and V. Trimble, A. I. P. (N. Y.).
\smallskip
\noindent Kent, S. M. 1986, AJ, 91, 1301.
\smallskip
\noindent Kent, S. M. 1987, AJ, 93, 816.
\smallskip
\noindent Kent, S. M., and Gunn, J. E. 1982, AJ, 87, 945.
\smallskip
\noindent Krumm, N., and Burstein, D. 1984, AJ, 89, 1319.
\smallskip
\noindent Liboff, R. L. 1990, Kinetic Theory, Prentice Hall, Englewood Cliffs,
N.J.
\smallskip
\noindent Mannheim, P. D. 1990, Gen.Relativ.Grav., 22, 289.
\smallskip
\noindent Mannheim, P. D. 1992, ApJ, 391, 429.
\smallskip
\noindent Mannheim, P. D. 1993a, Gen.Relativ.Grav., 25, 697.
\smallskip
\noindent Mannheim, P. D. 1993b, ApJ, 419, 150.
\smallskip
\noindent Mannheim, P. D. 1994, Foundations of Physics, 24, 487.
\smallskip
\noindent Mannheim, P. D., and Kazanas, D. 1989, ApJ, 342, 635.
\smallskip
\noindent Mannheim, P. D., and Kazanas, D. 1991, Phys.Rev.D, 44, 417.
\smallskip
\noindent Mannheim, P. D., and Kazanas, D. 1994, Gen.Relativ.Grav., 26, 337.
\smallskip
\noindent Milgrom, M. 1983a, ApJ, 270, 365.
\smallskip
\noindent Milgrom, M. 1983b, ApJ, 270, 371.
\smallskip
\noindent Milgrom, M. 1983c, ApJ, 270, 384.
\smallskip
\noindent Ostriker, J. P., and Peebles, P. J. E. 1973, ApJ, 186, 467.
\smallskip
\noindent Paresce, F., De Marchi, G., and Romaniello, M. 1995, ApJ, 440, 216.
\smallskip
\noindent Roberts, M. S., and Whitehurst, R. N. 1975, ApJ, 201, 327.
\smallskip
\noindent Rubin, V. C., and Ford, W. K. 1970, ApJ, 159, 379.
\smallskip
\noindent Rubin, V. C., Ford, W. K., and Thonnard, N. 1978, ApJ, 225, L107.
\smallskip
\noindent Rubin, V. C., Ford, W. K., and Thonnard, N. 1980, ApJ, 238, 471.
\smallskip
\noindent Rubin, V. C., Ford, W. K., Thonnard, N., and Burstein, D. 1982,
ApJ, 261, 439.
\smallskip
\noindent Rubin, V. C., Burstein, D., Ford, W. K., and Thonnard, N. 1985,
ApJ, 289, 81.
\smallskip
\noindent Sackett, P. D., Morrison, H. L., Harding, P., and Boroson, T. A.
1994,
Nature, 370, 441.
\smallskip
\noindent Sanders, R. H. 1990, A\&AR, 2, 1.
\smallskip
\noindent The, L. S., and White, S. D. M. 1986, AJ, 92, 1248.
\smallskip
\noindent Toomre, A. 1963, ApJ, 138, 385.
\smallskip
\noindent Tully, R. B., and Fisher, J. R. 1977, A\&A, 54, 661.
\smallskip
\noindent Udalski, A., Szymanski, M., Kaluzny, J., Kubiak, M., Krzeminski, W.,
Mateo, M., Preston, G. W., and Paczynski, B. 1993, Acta Astronomica, 43, 289.
\smallskip
\noindent Udalski, A., Szymanski, M., Kaluzny, J., Kubiak, M.,
Mateo, M., and Krzeminski, W. 1994, ApJ, 426, L69.
\smallskip
\noindent van Albada, T. S., Bahcall, J. N., Begeman, K. G., and Sancisi,
R. 1985, ApJ, 295, 305.
\smallskip
\noindent van Albada, T. S., and Sancisi, R.
1986, Phil.Trans.R.Soc.Lond., A, 320, 447.
\smallskip
\noindent  van der Kruit, P. C., and Searle L. 1981, A\&A, 95, 105.
\smallskip
\noindent Wevers, B. M. H. R., van der Kruit, P. C., and Allen, R. J. 1986,
A\&AS, 66, 505.
\smallskip
\noindent White, S. D. M., Navarro, J. F., Evrard, A. E., and Frenk, C. S.
1993,
Nature, 366, 429.
\vfill\eject
\noindent
{\bf Figure Captions}
\medskip
\noindent
Figure (1). The calculated rotational velocity curves associated with the
conformal
gravity potential $V(r)=-\beta c^2/r+ \gamma c^2r/2 $ for the four
representative
galaxies, the intermediate sized NGC 3198, the compact bright NGC 2903,
the large bright NGC 5907, and the dwarf irregular DDO 154 (at two possible
adopted distances). In each graph
the bars show the data points with their quoted errors, the full
curve shows the overall theoretical velocity prediction (in km/sec)
as a function of distance (in arc minutes) from the center of each galaxy,
while the two indicated dotted curves show the rotation curves that the
separate Newtonian and linear potentials would produce when integrated
over the luminous matter distribution of each galaxy. No dark matter is
assumed.
\smallskip
\noindent
Figure (2). The flattest possible rotation curve for a
thin exponential disk of stars each with conformal
gravity potential $V(r)=-\beta c^2/r+ \gamma c^2r/2 $
which is obtained
when the dimensionless
ratio $\eta$ takes the value 0.069. The full
curve shows the overall theoretical velocity prediction (in units of
$v/v_0$) as a function of distance (in units $R/R_0$),
while the two indicated dotted curves show the rotation curves that separate
Newtonian and linear potentials would produce. In the upper diagram the
rotation
curve is plotted out to 10 scale lengths to fully exhibit its flatness, while
in the lower diagram it is plotted out to 15 scale lengths to exhibit its
eventual asymptotic rise.
\smallskip
\noindent
Figure (3). The projected line of sight velocity data for the Coma cluster (as
binned by The and
White) plotted as function of impact parameter distance (in arc min) from the
center of the cluster.
\smallskip
\noindent
Figure (4). The Newtonian potential expectation for $<\sigma_p^2(R)>^{1/2}$ for
the generic modified Hubble profile matter distribution of Eq. (78) with
cut-off $R_M=20R_0$
in the pure isotropic, pure circular and pure radial
velocity cases. The velocity is normalized to $(N\beta_{gal} c^2/R_0)^{1/2}$
and the impact
parameter distance is plotted in units of the core radius $R_0$.
\smallskip
\noindent
Figure (5). The linear potential expectation for $<\sigma_p^2(R)>^{1/2}$ for
the generic modified Hubble profile matter distribution of Eq. (78) with
cut-off $R_M=20R_0$
in the pure isotropic, pure circular and pure radial
velocity cases. The velocity is normalized to $(N\gamma_{gal} c^2R_0)^{1/2}$
and the impact
parameter distance is plotted in units of the core radius $R_0$.
\smallskip
\noindent
Figure (6). The fractional amount of matter within a given volume of radius
$r$, and the
fractional amount of matter within a given surface of impact parameter $R$,
both
calculated for
the generic modified Hubble profile matter distribution of Eq. (78) with
cut-off $R_M=20R_0$.
The respective
distances ($r$ and $R$) are both plotted (on the same axis) in units of the
core radius $R_0$.
\smallskip
\noindent
Figure (7). The partial root mean square average projected line of sight virial
velocity
$\sigma_p(r_m)$
associated with Eq. (81) plotted as a function of radial distance from the
center of the cluster
in the Newtonian and linear cases for
the generic modified Hubble profile matter distribution of Eq. (78) with
cut-off $R_M=20R_0$.
The Newtonian potential velocity is normalized to $(N\beta_{gal}
c^2/R_0)^{1/2}$,
the linear potential velocity is normalized to $(N\gamma_{gal} c^2R_0)^{1/2}$,
and the radial distance is plotted in units of the core radius $R_0$.
\end